\documentclass[lettersize,journal]{IEEEtran}
\usepackage{amsmath,amsfonts}
\usepackage{array}
\usepackage[caption=false,font=scriptsize,labelfont=sf,textfont=sf]{subfig}
\usepackage{textcomp}
\usepackage{stfloats}
\usepackage{url}
\usepackage{verbatim}
\usepackage{graphicx}
\usepackage{cite}
\newtheorem{theorem}{Theorem}[section]
\usepackage{multirow}
\usepackage{booktabs} 
\usepackage{algorithm2e}
\usepackage{makecell}
\usepackage{tabularx}

\DeclareMathOperator{\atantwo}{atan2}
\AtBeginDocument{%
  \providecommand\BibTeX{{%
    \normalfont B\kern-0.5em{\scshape i\kern-0.25em b}\kern-0.8em\TeX}}}

\newtheorem{finding}{Finding}[section]
\SetKwInput{KwInput}{Input}                
\SetKwInput{KwOutput}{Output}              
\ifCLASSOPTIONcompsoc
  \usepackage[nocompress]{cite}
\else
  \usepackage{cite}
\fi
\begin{document}

\title{Frequency Bias Matters: Diving into Robust and Generalized Deep Image Forgery Detection}

\author{Chi~Liu,
        ~Tianqing~Zhu*,
        ~Wanlei~Zhou,~\IEEEmembership{Life Fellow,~IEEE,}
        ~Wei~Zhao,~\IEEEmembership{Fellow,~IEEE,}
\thanks{Chi Liu, Tianqing Zhu and Wanlei Zhou are with the Faculty of Data Science, City University of Macau, Macao SAR, China.  E-mail: chiliu@cityu.edu.mo; tqzhu@cityu.edu.mo; wlzhou@cityu.edu.mo}
\thanks{Wei Zhao is with Shenzhen Institute of Advanced Technology, Chinese Academy of Sciences, Shenzhen, China. \protect E-mail: zhao.wei@siat.ac.cn}
\thanks{*Tianqing Zhu is the corresponding author.}
\thanks{Manuscript received Jul .., 2024;}
\thanks{Acknowledgment: This work was supported by the National Natural Science Foundation of China (Grant No. 62402009), and the Science and Technology Development Fund of Macao under Grant 0013-2024-ITP1.}}

\markboth{Accepted for publication in IEEE Transactions on Dependable and Secure Computing}%
{Shell \MakeLowercase{\textit{et al.}}: Frequency bias matters: diving into robust and generalized deep image forgery detection}

\IEEEpubid{0000--0000/00\$00.00~\copyright~2021 IEEE}

\maketitle

\begin{abstract}
As deep image forgery powered by AI generative models, such as GANs, continues to challenge today's digital world, detecting AI-generated forgeries has become a vital security topic. Generalizability and robustness are two critical concerns of a forgery detector, determining its reliability when facing unknown GANs and noisy samples in an open world. Although many studies focus on improving these two properties, the root causes of these problems have not been fully explored, and it is unclear if there is a connection between them. Moreover, despite recent achievements in addressing these issues from image forensic or anti-forensic aspects, a universal method that can contribute to both sides simultaneously remains practically significant yet unavailable. In this paper, we provide a fundamental explanation of these problems from a frequency perspective. Our analysis reveals that the frequency bias of a DNN forgery detector is a possible cause of generalization and robustness issues. Based on this finding, we propose a two-step frequency alignment method to remove the frequency discrepancy between real and fake images, offering double-sided benefits: it can serve as a strong black-box attack against forgery detectors in the anti-forensic context or, conversely, as a universal defense to improve detector reliability in the forensic context. We also develop corresponding attack and defense implementations and demonstrate their effectiveness, as well as the effect of the frequency alignment method, in various experimental settings involving twelve detectors, eight forgery models, and five metrics.
\end{abstract}

\begin{IEEEkeywords}
Image forgery detection, generalization, robustness, frequency.
\end{IEEEkeywords}

\section{Introduction}\label{sec:introduction}
\IEEEPARstart{T}{he} recent advancements in deep generative models, particularly generative adversarial networks (GANs) \cite{goodfellow2020generative}, have remarkably improved automated image processing techniques. Alongside the success, deep face forgery technologies powered by cutting-edge generative models, such as DeepFake \cite{thies2016face2face}, are raising serious security concerns regarding individuals' safety \cite{nguyen2022deep,liu2020privacy}. 

As a response, research on countering forged face images has become a focus among security communities. One promising solution involves developing deep learning-based detectors capable of distinguishing AI-generated forged images from real ones \cite{nguyen2022deep, tyagi2023detailed}. Reliability is always a critical concern in developing forgery detectors, as it determines their applicability to broader, real-world scenarios. A detector's reliability is commonly assessed by two key properties: its generalization ability to detect forged images created by unknown GANs, and its robustness against arbitrary perturbation attacks \cite{wang2020cnn, he2021beyond, jeong2022frepgan}. 

Existing achievements on the generalization and robustness problems can be divided into forensic and anti-forensic directions. Current forensic studies generally focus on designing sophisticated detector networks or feature engineering methods to enhance these two properties \cite{chai2020makes,zhang2019detecting,wang2020cnn,bui2022repmix,jeong2022frepgan,he2021beyond,ojha2023towards,xu2023learning}. However, these efforts are incremental, outcome-driven, and case-by-case, which fail to provide a fundamental solution and will become increasingly laborious and challenging as the technologies behind forgery GANs and perturbation attacks continuously advance. Anti-forensic studies often conduct security analyses, sometimes devising novel attacks to expose a detector's vulnerability under various attacks \cite{hussain2021adversarial, neekhara2021adversarial, carlini2020evading, gandhi2020adversarial, fan2021deepfake, barni2019transferability, zhao2020effect}. These studies frequently conclude with empirical observations, underestimating the root causes of vulnerabilities, and they focus solely on robustness. Moreover, these attacks are often task-specific, requiring knowledge of the target detector, and are poorly transferable across different detectors.

Revisiting the above challenges from both forensic and anti-forensic sides, we argue that a critical concern is the lack of a high-level understanding that fundamentally explains why deep neural network (DNN)-based forgery detectors easily suffer from generalization and robustness issues despite their outstanding learning capabilities. It remains unclear whether there is an inner connection between generalizability and robustness in AI-generated forgery detection. This knowledge, intriguingly, could benefit both forensic and anti-forensic research simultaneously. For example, it might inspire a universal method to improve both the generalizability and robustness of a detector or facilitate the design of a novel attack capable of evading various detectors.
\IEEEpubidadjcol 
To address these challenges, we examine GAN-based forgery detection from a frequency perspective. Building on evidence that high-frequency bands capture subtle spectral artifacts and edge-level inconsistencies introduced by generation and post-processing \cite{dzanic2020fourier,frank2020leveraging,durall2020watch}, we provide a principled account of detector generalizability and robustness. Our analysis shows that frequency discrepancies between real and forged images induce a DNN \emph{frequency bias} that couples these two properties. This bias is largely driven by high-frequency components in training data, making detectors sensitive to high-band fluctuations. As a result, strongly biased detectors falter on unseen GANs and adversarial samples, which exhibit out-of-distribution high-frequency patterns.
\IEEEpubidadjcol
\begin{figure}
    \centering
    \includegraphics[width=0.48\textwidth]{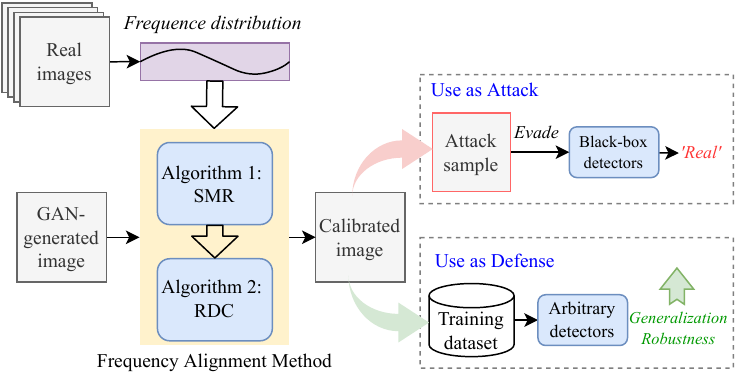}
    \caption{Overview of the proposed frequency alignment method and its different usages in attack and defense scenarios.}
    \label{fig:overview}
\end{figure}

Leveraging this insight, we propose a frequency alignment approach that narrows spectral gaps between real and forged images, benefiting both forensics and anti-forensics. As shown in Figure \ref{fig:overview}, our method combines a Spectral Magnitude Rescaling algorithm with a Reconstructive Dual-domain Calibration procedure to align fake-image spectra to those of real images in a coarse-to-fine manner. This yields two-sided gains from both forensic and anti-forensic viewpoints: as a strong black-box attack, frequency-aligned forgeries move inherently closer to real images, \emph{thus they can evade arbitrary detectors without accessing the target detector}; as a defense, frequency alignment reduces detector frequency bias, improving generalization and robustness without altering model architectures. We further present three defense implementations based on frequency-aligned samples, \emph{all architecture-agnostic and compatible with diverse detectors}.

The contributions of this paper are as follows:

\begin{itemize}
    \item We established a comprehensive, unified frequency analysis framework for AI-generated image detection. Through the analysis, we confirmed the frequency bias of DNN-based detectors, which can fundamentally explain several open problems related to the generalizability and robustness of DNN-based detectors. 
    \item We proposed a universal two-step frequency alignment method for refining AI-generated images by removing their frequency discrepancy from real images. The method can by applied to fake images created by diverse forgery models, including different GANs and different perturbation attacks. 
    \item The frequency alignment method can benefit the community from both forensic and anti-forensic sides. We proposed the corresponding attack and defense implementations, respectively, and verified the effects interactively in a wide range of settings. Ten baseline detectors, eight baseline forgery models, and five metrics are considered in the evaluation. 
\end{itemize}

\section{Related work}
\subsection{AI-generated image forgery detection}
\subsubsection{The forensic perspective}
\paragraph{Normal detectors}
Normal detectors aim for high detection accuracy when identifying AI-generated images in a known dataset. The recent technology can be divided into two mainstreams: image-domain detection and frequency-domain detection. 

Image-domain methods extract traces directly from pixels. Early work trained DNN classifiers to learn features end-to-end \cite{marra2018detection, tariq2018detecting, tyagi2023mininet}. Recent methods combine heuristic features or tailored learning schemes with DNNs. For instance, Nataraj et al. \cite{nataraj2019detecting} and Barni et al. \cite{barni2020cnn} used co-occurrence matrices across color channels. McCloskey et al. \cite{mccloskey2019detecting} exploited differences in color formation between cameras and GANs, building a detector based on saturated and underexposed pixels. Hu et al. \cite{hu2021exposing} showed that inconsistent corneal specular highlights in GAN faces are discriminative. Several works leverage GAN-specific “fingerprints,” analogous to camera fingerprints, extracted as pixel residuals \cite{marra2019gans} or encoded via DNNs from global representations \cite{yu2019attributing}. Others refine architectures or learning schemes, such as incremental learning to track new forgeries \cite{marra2019incremental} and lightweight self-attention modules for efficient fine-tuning with limited data \cite{jeon2020fdftnet}.

Frequency-domain methods learn from spectral representations, motivated by consistent frequency discrepancies between real and GAN images and across GANs \cite{dzanic2020fourier,frank2020leveraging,durall2020watch}. Images are transformed to spectra (e.g., Fourier, DCT), enabling detectors to capture these gaps. Simple classifiers, including shallow CNNs, perform well on DCT inputs \cite{frank2020leveraging}. Lightweight variants reduce 2D FFT magnitudes to 1D profiles \cite{dzanic2020fourier,durall2020watch}. Liu et al. \cite{liu2021spatial} found phase spectra exhibit stronger discrepancies than amplitudes and combined pixels with phase features. However, these frequency cues are fragile and easily perturbed \cite{durall2020watch, huang2020fakepolisher, jung2021spectral, dong2022think}, limiting robustness and cross-domain generalization.

A related branch targets complex forgeries involving post-processing (e.g., face alignment, rendering, compression). These methods follow similar principles, using image-, frequency-, or hybrid features \cite{ding2023dcu, guo2023hierarchical,sushir2024enhanced}. This paper focuses on end-to-end AI-generated face forgeries.

\paragraph{Generalized and robust detectors}
Beyond accuracy, recent work emphasizes generalization and robustness. Approaches include richer architectures, training strategies, and feature engineering. Chai et al. \cite{chai2020makes} analyzed semantic cues that generalize across GANs and proposed a patch-based classifier with limited receptive fields to emphasize local, transferable patterns. Zhang et al. \cite{zhang2019detecting} introduced AutoGAN to simulate common spectral artifacts for training generalized detectors. Data augmentations (e.g., JPEG, blur) can improve generalization \cite{wang2020cnn}. Bui et al. \cite{bui2022repmix} proposed representation mix-up and a novel loss for invariance to semantic changes and common transformations. Jeong et al. \cite{jeong2022frepgan} used frequency-level adversarial perturbations during training to suppress unstable GAN-specific artifacts. He et al. \cite{he2021beyond} leveraged re-synthesis residuals from a real-image pre-trained model as robust features. Ojha et al. \cite{ojha2023towards} showed fixed, large vision-language features generalize well, even without training. Xu et al. \cite{xu2023learning} exploited pairwise learning and complementary color-space representations.

\subsubsection{The anti-forensic perspective}
Anti-forensic methods probes an AI-generated forgery detector vulnerabilities via targeted attacks under white- and black-box settings, so as to evaluate the robustness of the forgery detector.

Most efforts use adversarial examples, adding imperceptible perturbations—often gradient-based—to cause misclassification. Classic attacks, including FGSM \cite{goodfellow2014explaining}, iterative FGSM \cite{kurakin2018adversarial}, CW $l_2$-norm \cite{carlini2017towards}, DeepFool \cite{moosavi2016deepfool}, and PGD \cite{madry2017towards}, have been applied to GAN-forgery detectors in both settings \cite{hussain2021adversarial, neekhara2021adversarial, carlini2020evading, gandhi2020adversarial, fan2021deepfake, barni2019transferability, zhao2020effect}. Further improvements include focusing perturbations on key forged regions \cite{liao2021imperceptible} and attacking in transformed color spaces to reduce perceptual degradation \cite{wang2021perception}.

Beyond adversarial noise, specialized reconstruction-based attacks re-synthesize forgeries to attenuate detectable traces. Examples include FakePolisher, which projects DeepFakes onto real-image manifolds to suppress spectral artifacts \cite{huang2020fakepolisher}; GANprintR, an autoencoder trained on real images to remove GAN fingerprints \cite{neves2020ganprintr}; adversarially trained re-synthesis to narrow real–fake gaps \cite{ding2021anti}; bidirectional conversion between GAN and natural faces \cite{peng2022bdc}; and TR-Net, which jointly removes multiple traces \cite{liu2022making}.

\subsection{Frequency analysis of DNNs' behavior}
Pioneering studies analyze DNNs via frequency, revealing biases across frequency bands and their impact on out-of-distribution generalization \cite{xu2019frequency, wang2020high, rahaman2019spectral}. Others examine adversarial robustness from a spectral viewpoint \cite{yin2019fourier}. However, frequency-level understanding of AI-generated image detection remains underestimated, where the insights from natural image classification do not directly transfer to AI-generated imagery which exhibit distinct frequency patterns. Moreover, the relationship between generalizability and robustness in AI-generated forgery detection remains insufficiently understood.

\section{Frequency analysis of forgery detectors}
\label{sec:fafd}
This section presents an empirical analysis of the generalization ability and robustness of AI-generated image forgery detection from the frequency perspective. AI-generated forgery detection is commonly formulated as a binary "real/fake" classification problem \cite{nguyen2022deep}. 
The generalization defines the cross-GAN detection ability of the detector, i.e., whether the detector can predict accurately facing test fake samples generated by unseen GANs not included in the training set $\mathbb{D}$. The robustness measures the reliability of the detector in detecting noisy fake samples manipulated by certain perturbation attacks. 


\subsection{Frequency Analysis tools}
\paragraph{Fourier transformation}
We adopt the 2D discrete Fourier transform (DFT) for image frequency analysis. Given an image $I \in \mathbb{R}^{M \times N}$, the frequency responses $\mathcal{F}(u, v)$ are computed as: 

\begin{equation}
\begin{aligned}
\mathcal{F}(I)(u,v)=\sum_{x=0}^{M-1} \sum_{y=0}^{N-1} I(x, y) \cdot e^{-2 \pi i \cdot \frac{u x}{M}} e^{-2 \pi i \cdot \frac{v y}{N}}\\ \text { for } x=0,1, \ldots, M-1, \quad y=0,1, \ldots, N-1
\end{aligned}
\end{equation}
This transform is reversible and we denote the inverse DFT that transforms spectrum back to image as $\mathcal{F}^{-1}(\cdot)$. The DFT spectrum is typically visualized in a form of center-shifted magnitude heatmap, where lower frequency components are closer to the center of the spectrum while higher frequency components are farther from. 

\begin{figure}
    \centering
    \includegraphics[width=0.48\textwidth]{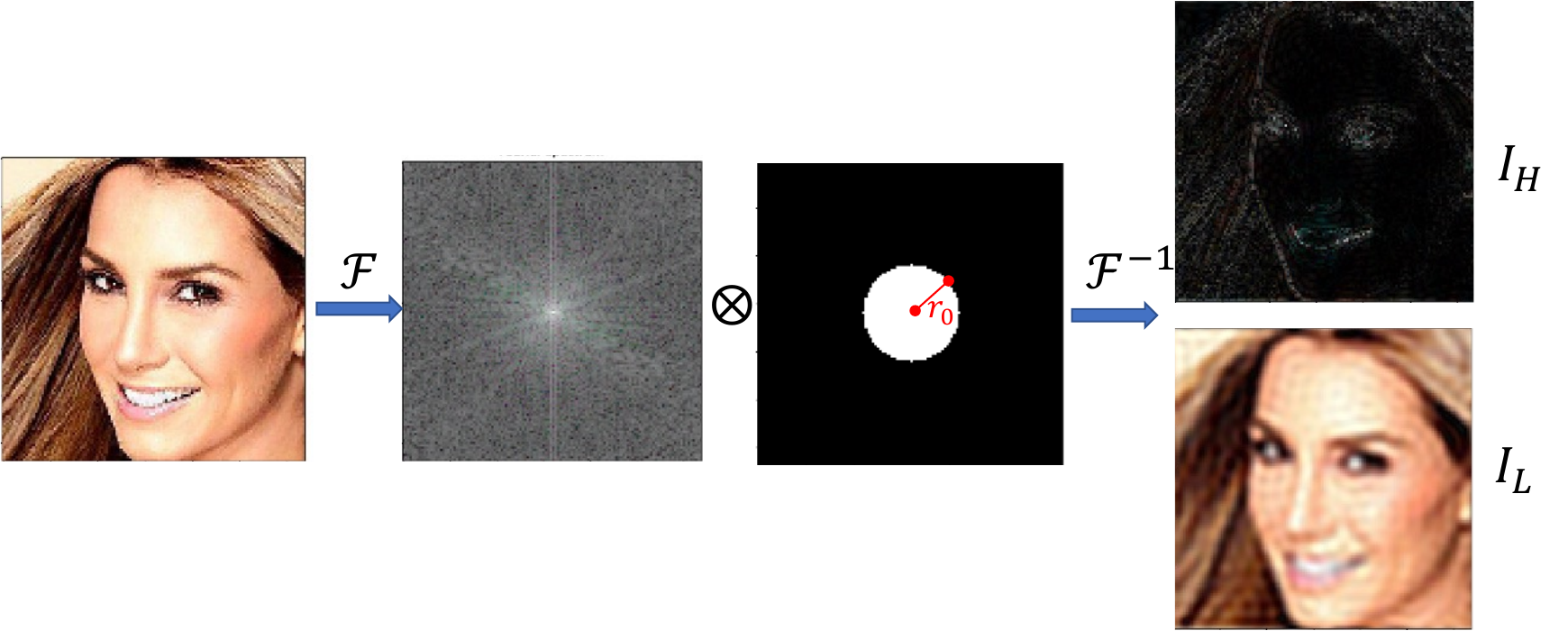}
    \caption{The process of frequency decomposition. A circular mask-based ideal binary filter is applied to the center-shifted DFT spectrum of the image to decompose it into high- and low-frequency components. }
    \label{fig:decomposition}
\end{figure}

\paragraph{Frequency decomposition}
Our frequency analysis requires to decompose an image $I$ into the low-frequency and high-frequency components, i.e., $I=\{I_{L}, I_{H}\}$. This can be done by applying a filter to the center-shifted DFT spectrum of the image. We use a circular mask-based ideal filter $\tau(r_{0})$ with a predefined radius $r_{0}$ for decomposition, denoted as : 
\begin{equation}
\label{eq:decom}
\left\{
\begin{aligned}
I_{L} &= \mathcal{F}^{-1}\left(\tau(r_{0}) \otimes \mathcal{F}(I) \right)\\
I_{H} &= \mathcal{F}^{-1}\left((1-\tau(r_{0})) \otimes \mathcal{F}(I) \right)
    \end{aligned}\right.
\end{equation}
where $\otimes$ is element-wise multiplication and each element in $\tau(r_{0})$ is defined as:
\begin{equation}
\begin{aligned}
    \tau(r_{0})_{u,v} = \left\{\begin{aligned} & 1, && \text{if} \quad \sqrt{(u-u_0)^2+(v-v_0)^2}\leq r_{0} ,\\
    &0, && \text{otherwise}, 
    \end{aligned} \right. \\
\text { for } u=0,1, \ldots, M/2-1, \quad v=0,1, \ldots, N/2-1    
\end{aligned}
\end{equation}
where $(u_0, v_0)$ is the coordinate of the centroid. Figure \ref{fig:decomposition} shows the decomposition of an image example with a certain radius. 

\paragraph{Frequency distribution}
In order to straightforward observe the frequency discrepancy in a statistical view, we estimate the frequency distribution of a DFT spectrum. Given the rotation invariance of a the center-shifted DFT spectrum, the frequency distribution can be represented as a one-dimensional profile via azimuthally integrating the spectral magnitudes over the radial frequencies $\theta$ \cite{durall2020watch}. Assuming a square image $I \in \mathbb{R}^{N \times N}$, its one-dimensional profile is:

\begin{equation}
\label{eq:fd}
\operatorname{FD}\left(r_{k}\right)=C_0\int_{0}^{2 \pi}|\mathcal{F}(r_{k}, \theta)| \mathrm{~d} \theta \quad \text{for}\quad k=0,1,...,N/2-1, 
\end{equation}
where $C_0$ is a normalization constant, $(r_{k}, \theta)$ is the polar coordinate transformed from $(u,v)$: $r_{k}=\sqrt{u^{2}+v^{2}}$, $\theta=\atantwo\left(v, u\right)$.
For ease we normalize $r_k$ into the range of $[0, 1]$ using the factor $\frac{1}{\sqrt{\frac{1}{2} N^2}}$, and use a log-scaled spectrum instead of the raw spectrum.

\begin{figure*}
     \centering
     \subfloat[\rmfamily{The average DFT spectra. Unlike real images with limited magnitude in higher-frequency bands, both AI-generated and perturbation samples introduce spectral artifacts, yielding abnormal higher-frequency responses. Notably, StyleGAN and FGSM samples exhibit clear spectral checkerboard patterns.}\label{fig:spectral_dis}]{\includegraphics[width=0.48\textwidth]{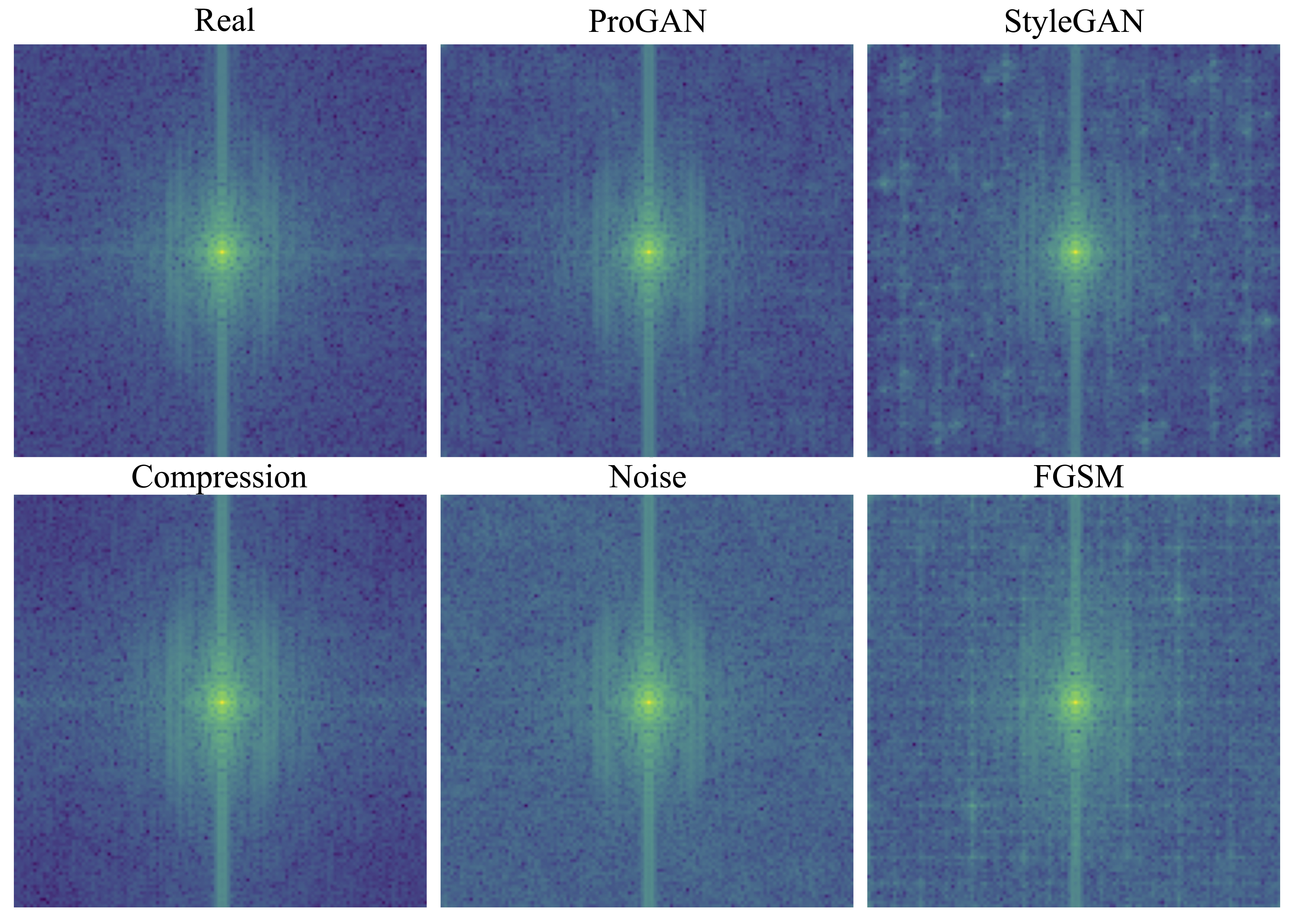}}\hfill
     \subfloat[\rmfamily{The spectral distributional gaps between real and each forgery type confirm the specific frequency discrepancy.}\label{fig:distributional_dis}]{\includegraphics[width=0.5\textwidth]{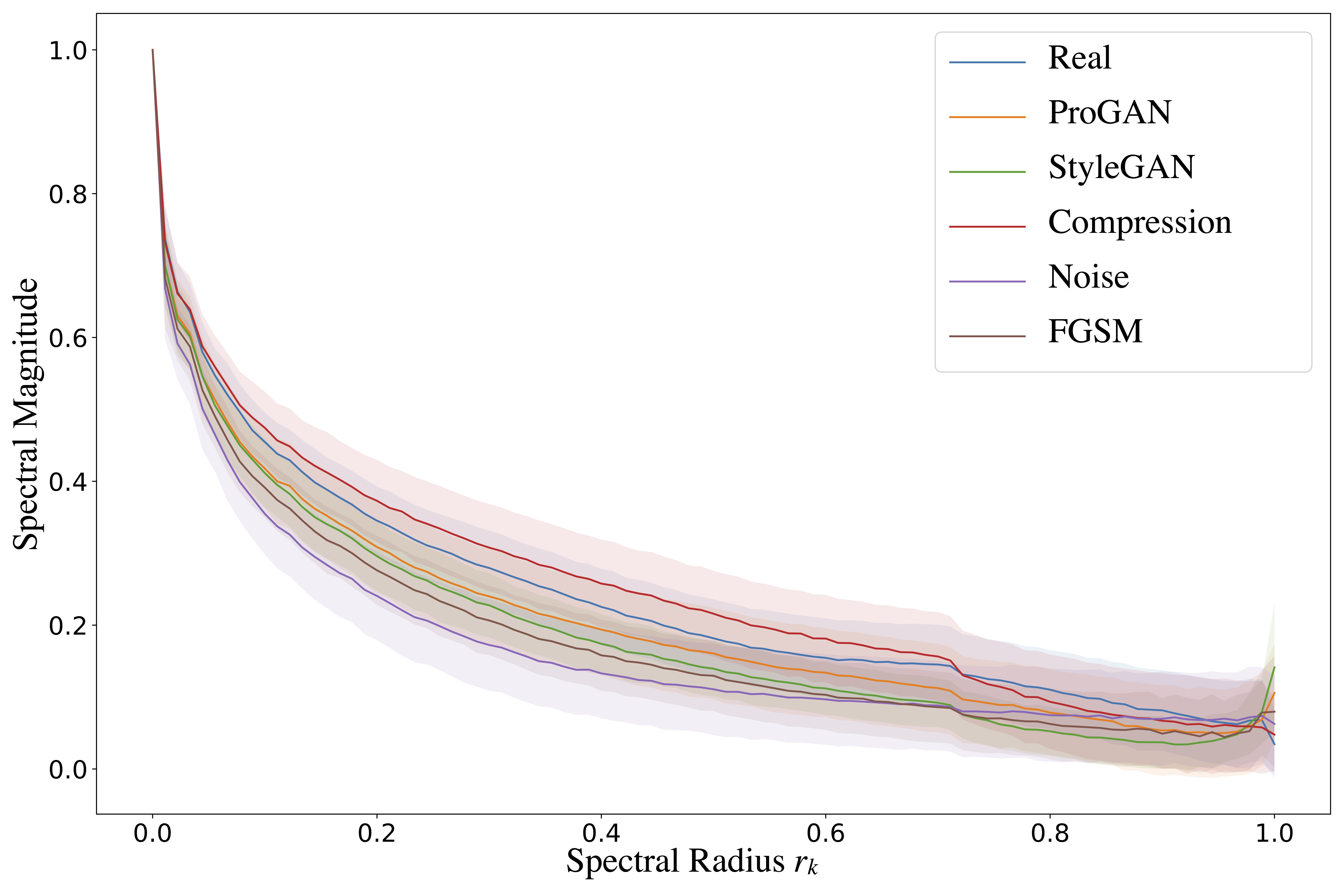}}\hfill
        \caption{Visualizations of frequency discrepancies in real images, AI-generated images, and perturbation examples crafted from ProGAN images.}
        \label{fig:three graphs}
\end{figure*}

\subsection{Visualization of frequency discrepancy}
\label{sec:vfd}
We first provide the spectral visualizations of different forgery models, including different GANs and different perturbations, to empirically confirm the existence of the frequency discrepancies and how they present. 
The selected forgery models include two popular GANs (ProGAN \cite{karras2017progressive} and StyleGAN \cite{karras2019style}) trained on the human face dataset CelebA, and three representative perturbations crafted on ProGAN samples (Compression, Noising, and an adversarial attack FGSM ($\epsilon=4/255$)). 

Figure \ref{fig:spectral_dis} depicts the average DFT spectra of various forgery patterns. The average frequency distributions, as computed by Eq \ref{eq:fd}, are shown in Figure \ref{fig:distributional_dis}. Combining the two figures, the spectral discrepancies between real and fake images are clearly observed, along with two key findings: 1) Each forgery model has its specific frequency pattern, resulting in a unique spectral discrepancy from real images; 2) The spectral discrepancies generally become larger in higher frequency components, e.g., $r_{k} > 0.1$ for ProGAN. Notably, the findings hold for both AI-generated samples and perturbed samples, implying a potential theoretical connection between generalization ability and robustness from the frequency standpoint. 


\subsection{Frequency bias of detectors}
Now we seek to establish a unified explanation of the generalization and robustness problems of forgery detectors with the following frequency bias hypothesis:

\begin{finding}[The frequency bias of forgery detectors]
\label{pythagorean}
A CNN detector easily overfits the specific frequency discrepancy between the forgery images and real images in the training set, and thus fails to detect test forgery samples with a distinct frequency pattern.
\end{finding}

The frequency bias hypothesis can explain both the issues of generalization and robustness simultaneously. This is because when dealing with unseen forgery samples, whether they are produced by a different GAN or manipulated through a perturbation attack, they will possess a specific frequency pattern different from those found in the training set. Hence, if the hypothesis holds, a biased detector that has overfitted the frequency patterns in the training dataset is likely to struggle when identifying unseen forgery models with different frequency patterns. We offer two pieces of evidence to validate this hypothesis.

\textbf{Validation 1:} According to the observation of the high-frequency distributional characteristic of spectral discrepancy in Section \ref{sec:vfd}, we aim to evaluate the responses of forgery detectors to different frequency components. Concretely, we decompose images into a set of pairs of $I=\{I_{L}, I_{H}\}_{r_0} $ by changing $r_0$ following Eq \ref{eq:decom}. Then we discard the high-frequency component $I_{H}$ where the spectral discrepancies are likely to concentrate upon, and train and test the detectors with only the low-frequency components $I_{L}$. Four widely-used CNN-based forgery detector backbones, ResNet18 \cite{he2016deep}, DenseNet\cite{huang2017densely}, Xception \cite{chollet2017xception} and EfficientNet\cite{tan2019efficientnet}, are evaluated. All detectors are trained with real and clean ProGAN images and tested on different forgery models, including two GANs (ProGAN and StyleGAN), and three perturbations images (JPEG Compression, Gaussian Noise and the adversarial perturbation FGSM). The real images are sampled from the CelebA dataset, and the two GANs are pretrained with the same dataset to avoid domain shifts. The perturbations are crafted on the ProGAN test images. 

\begin{figure*}
    \centering
    \includegraphics[width=0.9\textwidth]{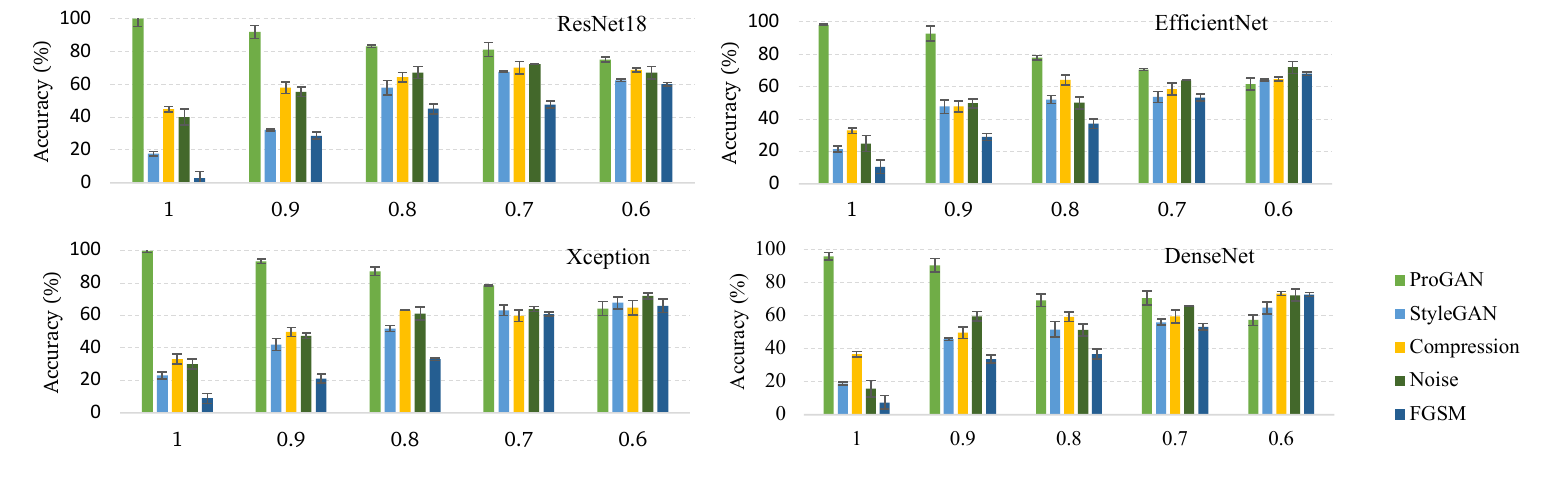}
    \caption{The generalization and robustness of four DNN detectors trained and tested on different frequency bands.}
    \label{fig:diff_bands}
\end{figure*}

Figure \ref{fig:diff_bands} shows the results. When evaluating the raw images with full frequency information (i.e., no filter), the intra-dataset tests on the same forgery model ProGAN achieve high accuracy, while the generalization to StyleGAN and the robustness against perturbations are poor. In the low-frequency groups, with decreasing the radius of the filter, which means more high-frequency components are excluded, the intra-dataset performances drop unsurprisingly due to information loss. However, the generalization and robustness increase significantly for both ResNet18 and Xception, which means the detectors behave more stably after reducing the reliance on high-frequency discrepancy. The results confirm the frequency bias. 

\textbf{Validation 2:} We further verify the hypothesis with integrating the Frequency Principle Theory of CNN classifiers. 

\begin{theorem}[Frequency Principle Theory of CNN]
DNNs often fit target functions from low to high frequencies during the training process \cite{xu2019frequency}.
\end{theorem} 

The theory describes a CNN classifier's tendency to first pick up low-frequency information and then overfit high-frequency information when learning natural images \cite{wang2020high}. Applying the theory to forgery detectors, it can be deduced that detectors will exhibit a more severe frequency bias as training progresses. This is because frequency discrepancies primarily occur in higher frequency components which are mostly captured in later training phases.  As a result, by evaluating the performance of the same detector at varying degrees of convergence, the influence of frequency bias can be verified.

\begin{figure}
    \centering
    \includegraphics[width=0.5\textwidth]{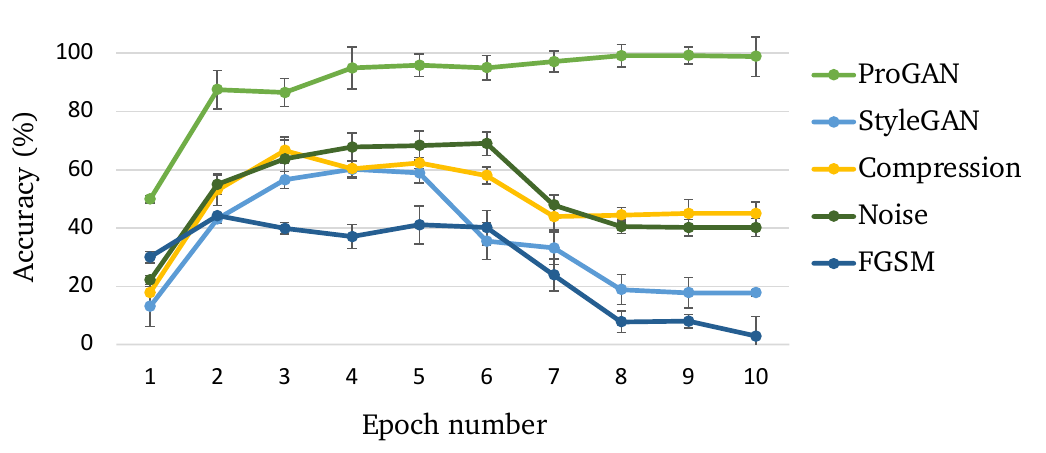}
    \caption{The generalization and robustness of the same DNN detector picked at different training epochs.}
    \label{fig:diff_epochs}
\end{figure}

To this end, we train a shallow CNN forgery detector using real and clean ProGAN images and test it with all forgery types at the end of each training epoch. Each epoch represents a certain convergence degree ranging from underfitting to overfitting. Figure \ref{fig:diff_epochs} shows the results. Before the detector converges (i.e., epoch$\leq6$), its test performances on unseen GANs or perturbations continue to improve. However, in later epochs, when the detector overfits more high-frequency information, the generalization ability and robustness both deteriorate remarkably. The outcomes again confirm the frequency bias.

\subsection{Insights from the analysis}
The frequency analysis of forgery detectors confirms the finding of frequency bias, a possible factor affecting the generalizability and robustness of forgery detectors and intrinsically connecting the two properties. Becasue the frequency bias is principally associated with the higher-frequency components of the training images which are simple-to-detect yet unstable features, the detectors will become sensitive to any changes in high-frequency bands. As a result, a detector with significant frequency bias will struggle to detect unknown GAN samples or noisy samples, because both of them manifest a different high-frequency pattern that is outside the frequency distribution of the training dataset. 

The finding of frequency bias motivates us to rethink the forgery detection problem: Considering the negative impacts of high-frequency discrepancy between images from different sources, if we can develop a method to reduce and even remove the in-between frequency discrepancies, it may benefit forgery detection from both image forensics and anti-forensics aspects:

\textbf{The anti-forensic aspect:} This method can be used directly as a strong black-box attack to evade forgery detectors. Unlike previous attacks that fool detectors by perturbing the frequency distribution of the original fake images, removing the frequency discrepancy of fake images makes them intrinsically closer to real ones, not only in vision perception but also from the frequency distribution aspect. Therefore, these manipulated fake images can serve as evasion attack samples without the need to access and the knowledge of the detector, and intuitively, they will possess better attack transferability across different detectors. 

\textbf{The forensic aspect:} The method can also be used to improve detectors' generalization and robustness, by training or fine-tuning detectors with manipulated fake images. This is because after the high-frequency discrepancies have been removed, the detectors retrained with these samples will become less dependent on unstable frequency patterns, reducing the frequency bias and alternatively focusing on learning more generic and robust features.

\section{The Frequency Alignment Method}
In this section, we propose the Frequency Alignment Method to eliminate the frequency discrepancy between real and fake images by aligning their frequency distributions. The frequency alignment process is a many-to-one mapping, aiming to calibrate all possible frequency distributions of fake images to the real ones. This method performs solely on the images and is detector-independent, without the need to access and the knowledge of the detector. 

\subsection{Problem formulation}
The Frequency Alignment problem can be formally formulated as follows:

Let $\mathbb{I}^{+}$, $\mathbb{I}^{-}$ be the original real and fake image datasets, respectively. We want a function $F: I^{*} = F(I^{-})$ that can modify a given fake sample $I^{-} \in \mathbb{I}^{-}$ to $I^{*}$ with satisfying the following goal:
\begin{equation}
\label{eq:prob_form}
    \min D(q({I}^{*}) || p({I}^{+})), \quad s.t. \quad \forall I^{-} \in \mathbb{I}^{-}, ||I^{-} - I^{*}|| \leq \epsilon
\end{equation}
where $q({I}^{*})$ and $p({I}^{+})$ indicate the frequency distributions of fake and real samples, respectively, and $D()$ is the divergence measurement. The constraint term ensures that the modification of the original fake sample by $F$ is small enough so that no perceptual image quality degradation is caused. 

To solve the problem, we propose a two-step method to achieve a coarse-to-fine alignment. The first step is called Spectral Magnitude Rescaling (SMR). We rescale the spectral magnitudes of fake samples based on the estimated fitting function of real images' frequency distribution. The second step is called Reconstructive Dual-domain Calibration (RDC), where a denoising auto-encoder is first learned with only real images to model both the pixel and frequency distributions of real images. Then the rescaled fake samples generated by Step 1 are reconstructed by the auto-encoder with a dual-domain calibration to real images in the latent feature space. 

\subsection{Spectral Magnitude Rescaling}
The SMR algorithm aims to reduce the high-frequency gap between real and fake samples by rescaling fake samples' spectral magnitudes. The rescaling factor is adaptively computed at each frequency band according to the ratio of the empirical frequency distributions of real and fake images. To this end, we need to model the frequency distribution with an estimated parametric equation. As previous studies have pointed out that the spectra of natural images distribute following a power law \cite{van1996modelling}, the expectation with respect to the frequency distribution can be modeled using a power law function: 
\begin{equation}
    \mathbb{E}({FD}(r_k)) \approx a \cdot r_k ^{b}
\end{equation}
where the parameter $a$ represents the spectral magnitude at the position $r_k$, and $b$ represents the decay rate of the spectrum. The two parameters can be estimated by fitting the power law function with a number of images' one-dimensional spectral profiles $FD(r_k)$. Then, the spectrum of a given fake sample $I^{-}$ can be rescaled as follows: 

\begin{equation}
\label{eq:simple_smr}
    \hat{\mathcal{F}}(I^{-})(r_{k}, \theta) = \mathcal{F}(I^{-})(r_{k}, \theta) \left[\frac{a^{+}}{a^{-}}(r_{k})^{b^{+}-b^{-}}\right]
\end{equation}
where $(a^{+}, b^{+})$ and $(a^{-}, b^{-})$ are the parameters estimated from real images and fake images, respectively. 

However, two practical challenges remain. Firstly, the visual content of the given fake sample, including facial details (such as profile, direction, and size), backgrounds, and color information, may differ significantly from the images used for fitting. This discrepancy can result in substantial visual distortions in the processed fake sample. Secondly, since the frequency discrepancy is predominantly found in high-frequency components, it is preferable to perform rescaling specifically on these high-frequency bands to suppress visual artifacts and reduce computational overhead.

\begin{figure*}
    \centering
    \includegraphics[width=0.9\textwidth]{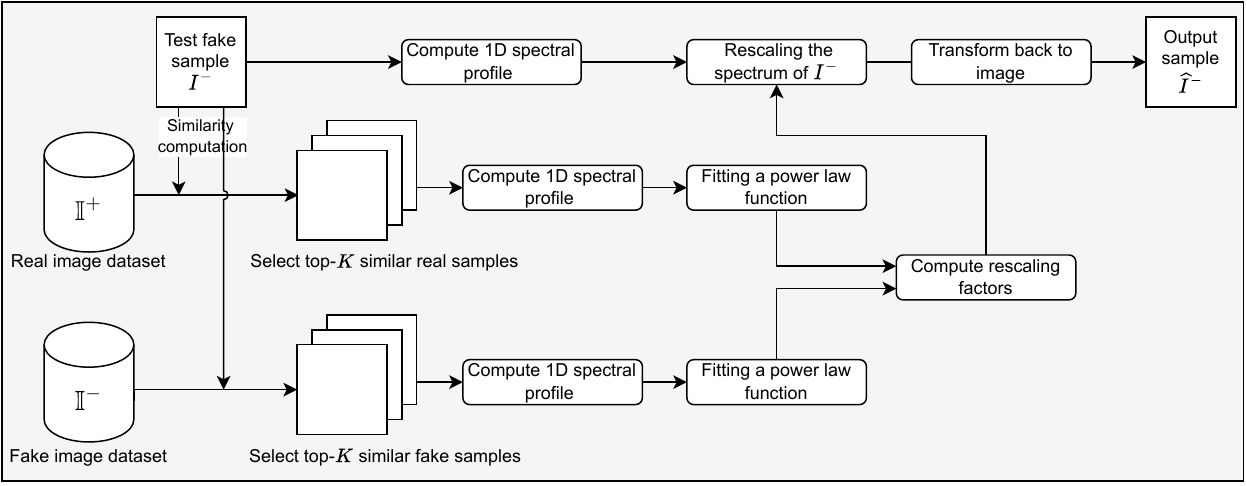}
    \caption{The workflow of the Spectral Magnitude Rescaling algorithm.}
    \label{fig:smr_workflow}
\end{figure*}

To overcome the above challenges, we have two improvements to the algorithm. First, instead of randomly sampling image samples for fitting the function, we retrieve top-$K$ similar samples that are visually close to the given fake sample from the real and fake image datasets individually. The retrieval is based on the Structural Similarity Index (SSIM) score. Second, we impose a threshold and smoothing factor to adjust the rescaling function in Eq.\ref{eq:simple_smr}:
\begin{equation}
\label{eq:final_smr}
\begin{aligned}
    \hat{\mathcal{F}}(I^{-})(r_{k}, \theta) = & \mathcal{F}(I^{-})(r_{k}, \theta) \left[1+\left(\frac{a^{+}}{a^{-}}(r_{k})^{b^{+}-b^{-}}-1\right) S(r_{k})\right],\\
    S(r_{k})= & \left\{
            \begin{aligned}
            & \frac{\mathrm{1} }{\mathrm{1} + e^{-(r_{k}-r_{T})}}, & r_{k}\geq r_{T}, \\
            & 0, & r_{k} < r_{T},
            \end{aligned}
            \right.
    \end{aligned}
\end{equation}
where $r_{T}$ defines a fixed threshold frequency band above which the rescaling is performed to enforce the low-frequency bands unaffected, $S(r_{k})$ is a sigmoid function when $r_{k}\geq r_{T}$ to smooth the rescaling. The entire SMR algorithm is shown in Algorithm \ref{al:smr} and the workflow is shown in Figure \ref{fig:smr_workflow}:  

\RestyleAlgo{ruled}
\SetKwComment{Comment}{/* }{ */}
\begin{algorithm}
\caption{Spectral Magnitude Rescaling}
\KwInput{The real image dataset $\mathbb{I}^{+}$; The fake image dataset $\mathbb{I}^{-}$; Sampling number $K$; Frequency threshold $r_{T}$; A given fake sample $I^{-}$; }
\KwResult{The spectrum-rescaled fake sample $\hat{I}^{-}$}
 1. Retrieving the $K$ samples most similar to $I^{-}$ from $\mathbb{I}^{+}$ and $\mathbb{I}^{-}$ independently\;
 2. Computing the 1D spectral profile $FD(r_k)$ for all selected samples \Comment*[r]{following Eq. \ref{eq:fd}}
 3. Estimating the parameters $(a^{+}, b^{+})$ and $(a^{-}, b^{-})$ by fitting a power law function on the sampled real and fake samples, respectively\;
 4. Transforming $I^{-}$ to its spectrum $\mathcal{F}(I^{-})$\;
 5. Rescaling $\mathcal{F}(I^{-})$ to $\hat{\mathcal{F}}(I^{-})$ \Comment*[r]{following Eq. \ref{eq:final_smr}}
 6. Transform $\hat{\mathcal{F}}(I^{-})$ back to the image domain: $\hat{I}^{-}$ = $\mathcal{F}^{-1}\left(\hat{\mathcal{F}}(I^{-})\right)$.
 \label{al:smr}
\end{algorithm}

\subsection{Reconstructive Dual-domain Calibration}
Although the SMR algorithm can reduce the high-frequency gap between real and fake samples, it will still remain several high-frequency artifacts. The reasons include that the visual contents of the real and fake samples selected for fitting do not exactly match, and the estimation of the fitting function is an empirical approximation. In order to further align the frequency patterns while satisfying the constraint of visual quality in Eq. \ref{eq:prob_form}, a more fine-grained calibration is needed. We propose the Reconstructive Dual-domain Calibration (RDC) algorithm. The key idea is to simulate both the pixel and frequency distributions of real images via a learnable model, and then use the model to calibrate the fake images resulting from the SMR algorithm in both the image and frequency domains. 

\subsubsection{Self-supervised denoising} 
\begin{figure*}
    \centering
    \includegraphics[width=0.9\textwidth]{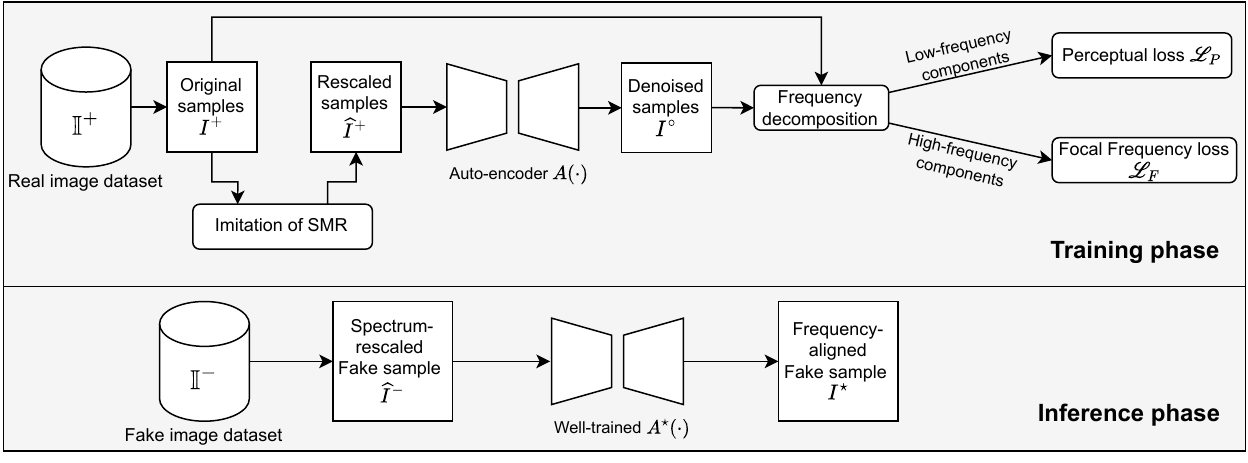}
    \caption{The workflow of the Reconstructive Dual-domain Calibration algorithm. In the training phase, a self-supervised denoising auto-encoder $A$ is trained with only real images, learning to model the distribution of real images in both the image and frequency domains. In the inference phase, the well-trained $A^{*}$ is applied to fake samples to calibrate the frequency patterns.}
    \label{fig:rdc_workflow}
\end{figure*}

We formulate the simulation of real images as a learning-based denoising process, i.e., try to reconstruct the original real image from its noised version by an auto-encoder $A(\cdot)$. As shown in Figure \ref{fig:rdc_workflow}, the auto-encoder is trained with the real image dataset $\mathbb{I}^{+}$ only. The correct pixel and high-frequency distributions of real images are then captured by the auto-encoder through reconstruction learning. In the inference phase, the well-trained $A^{*}(\cdot)$ is applied to reconstruct a given fake sample. The dual-domain calibration is completed in the latent feature space formed by $A^{*}(\cdot)$.  

To ensure an accurate calibration from the spectrum-rescaled fake samples to the real images, the noised real images, i.e., the inputs of $A(\cdot)$, should be initialized to a similar pattern as the spectrum-rescaled fake samples. However, the SMR algorithm cannot be applied directly to real images because real images themselves are the ground-truth reference for power law fitting. As an alternative, we propose an approximation method to imitate the effect of SMR. Given a real image $I^{+}$, we compute its noised version $\hat{I}^{+}$ as follows:

\begin{equation}
\begin{aligned}
    \hat{\mathcal{F}}(I^{+})(r_{k}, \theta) & = \mathcal{F}(I^{+})(r_{k}, \theta) \left[1+\left(a^{\prime}(r_{k})^{b^{\prime}}-1\right)S(r_{k})\right],\\
    \hat{I}^{+} & = \mathcal{F}^{-1}\left(\hat{\mathcal{F}}(I^{+})\right).
\end{aligned}
\end{equation}
where $S(r_{k})$ is the same as in Eq. \ref{eq:final_smr}. $a^{\prime}$ and $b^{\prime}$ are randomly sampled from $[1/2, 2]$ and $[-4, 4]$ respectively to simulate the rescaling factor in Eq. \ref{eq:final_smr}. Then, $A(\cdot)$ can be trained in a self-supervised reconstruction task, as shown in Figure \ref{fig:rdc_workflow}. 

\subsubsection{Network and Losses} We use a U-shaped encoder–decoder architecture \cite{ronneberger2015u} as the backbone of \(A(\cdot)\) due to its strong image reconstruction capability and efficient training. Our U-Net has four blocks in both the encoder and decoder. The encoder stacks \(3 \times 3\) convolutions with \(2 \times 2\) max pooling to extract multi-scale features while reducing spatial resolution. The decoder symmetrically upsamples to recover resolution. Skip connections link each encoder block to its corresponding decoder block, concatenating features to preserve spatial detail and strengthen representations for reconstruction.

The loss function supervises $A(\cdot)$ to reconstruct $I^{+}$ from $\hat{I}^{+}$. For our dual-domain reconstruction task, the conventional pixel-to-pixel reconstruction loss, $||I^{+} - A(\hat{I}^{+})||$, is impractical. As the majority of pixel information in a natural image is associated with low-frequency bands, the pixel-to-pixel loss can easily lead to a suboptimum that overfits the low-frequency component. Moreover, it fails to solve the issue of spectral artifacts caused by upsampling \cite{durall2020watch}. We approach this problem by decomposing images into low- and high-frequency components and addressing each component separately, detailed as follows: 

Given the original sample $I^{+}$ and its reconstructed version $I^{\circ} = A(\hat{I}^{+})$, following Eq. \ref{eq:decom}, we decompose them into $(I^{+}_L, I^{+}_H)_r$ and $(I^{\circ}_L, I^{\circ}_H)_r$ respectively, with a random radius threshold $r$. With regard to the low-frequency components, we compute the perceptual loss \cite{johnson2016perceptual} to measure the pixel similarity: 

\begin{equation}
\mathcal{L}_{P}= \frac{1}{K}\sum_{k=0}^{K-1}\left\|\operatorname{VGG}_{k}(I^{+}_L)-\operatorname{VGG}_{k}(I^{\circ}_L))\right\|,
\end{equation}
where $\operatorname{VGG}_{k}(\cdot)$ is the respective feature obtained by the $k$-th convolutional layer of a total of $K$ convolutional layers within a pre-trained VGG classification network. The perceptual loss can better recover the low-frequency visual details that correlate with the human visual system compared with the pixel-to-pixel loss. For the high-frequency components, we transform them into DFT spectra and compute the focal frequency loss \cite{jiang2021focal} to measure the frequency similarity:
\begin{equation}
\begin{aligned}
&\mathcal{L}_{F}=\frac{1}{M N} \sum_{u=0}^{M-1} \sum_{v=0}^{N-1} w(u, v)\left|\mathcal{F}(I^{+}_H)(u, v)-\mathcal{F}(I^{\circ}_H)(u, v)\right|^2 \\
&w(u, v)=\left|\mathcal{F}(I^{+}_H)(u, v)-\mathcal{F}(I^{\circ}_H)(u, v)\right|
\end{aligned}
\end{equation}
where $w(u, v)$ is a self-adaptive weight to force the model to focus more on higher frequency. The final training objective function is:
\begin{equation}
\min \mathcal{L} = \mathcal{L}_{P} + \lambda \mathcal{L}_{F},
\end{equation}

In the inference phase, the optimal model $A^{*}(\cdot)$ is applied to the spectrum-rescaled fake samples to create the final frequency-aligned samples with dual-domain calibration, denoted as:
\begin{equation}
    I^{*} = A^{*}(\hat{I}^{-})
\end{equation}

\subsection{Compared with other methods}
\label{sec:comparison}
\textbf{Compared with low-pass filter.} The low-pass filter is a straightforward way to remove the frequency discrepancies, given that the discrepancies heavily rely on high-frequency components. However, in most cases, the filter has a fixed kernel that fails to support sample-specific alignments. Also, filtering will cause unnecessary loss of high-frequency information. In contrast, our method involves a coarse-to-fine alignment based on learning the frequency pattern of real images, which is more flexible and smoother and can perfectly preserve the full-band information.  

\noindent \textbf{Compared with frequency regularization.} Some recent studies have proposed imposing an additional frequency regularization loss \cite{durall2020watch} or frequency discriminator \cite{jung2021spectral, chen2021ssd} on the source GAN to suppress its frequency distortion during training. Unlike our algorithm, which is post-processing and applicable to various forgery models, these methods only work for one specific GAN, require retraining the source GAN, and cannot be used to align forgery samples post-processed by perturbation models. 

\noindent \textbf{Compared with adversarial learning.} Another typical idea recently proposed is to train a model to directly reconstruct the fake samples via adversarial learning \cite{ding2021anti, liu2022making}. Alongside the reconstruction generator, a discriminator is needed to distinguish the reconstructed fake images from real ones in the frequency domain during training. Compared with our RDC algorithm trained only on real images, this kind of methods requires a large number of fake samples from various forgery models for training, which is hard to acquire in practice. Moreover, the discriminator will also suffer from the frequency bias \cite{xu2019frequency, wang2020high}, resulting in lower visual quality and alignment precision, as confirmed in our experiments.

\subsection{Attack and defense implementations}
\label{sec:def_imple}
As discussed earlier, the frequency alignment method enables both the anti-forensic (attack) and forensic (defense) usages. The implementations are as follows:

\noindent \textbf{Attack implementation.} Given an arbitrary victim detector $\mathcal{C}$, we perform the frequency alignment method to modify a fake sample $I^{-}$ into $I^{*}$. The modified sample $I^{*}$ is much more realistic and can directly serve as an attack sample to evade the detection by $\mathcal{C}$. Notably, as an attack, our method is detector-independent, requiring zero knowledge of $\mathcal{C}$. Thus, it works well for challenging black-box scenarios and has good cross-detector transferability.  

\noindent \textbf{Defense implementation.} The frequency alignment method can be used to improve the detector's generalization ability and robustness by forcing the detector to mine more generic frequency-irrelevant features while reducing frequency bias. We propose three implementation methods:

(1) Implementation as a simple data augmentation in the training phase. We set the probability of a training sample being modified by our method to $0.5$.

(2) Implementation as a pre-processing procedure in both the training and inference phases of the detector. 

(3) Hybrid implementation: We pre-process all training and test samples with the frequency alignment method, and also employ a mix-up augmentation with a probability of $0.5$. The mix-up augmentation is denoted as: 
    \begin{equation}
        \left\{
        \begin{aligned}
            I^{+}_{aug} &= I^{+} + \delta|I^{-} - I^{*}| \\
            I^{-}_{aug} &= I^{*} + \delta|I^{-} - I^{*}|
        \end{aligned}
        \right.
    \end{equation}
The key idea is adding the residual $|I^{-} - I^{*}|$ to the raw training samples to create hard learning samples. $\delta\sim \mathcal{N}(0,1)$ is used to scale the residual to enable various degrees of hardness. 
    
Note that all the defense methods are free of modifying the detector's network and are therefore compatible with various detectors.

\section{Experiments}
In the experiments we aim to

(1) evaluate the performances of the proposed frequency alignment method implemented as attack and defense separately, both in a wide range of settings; and 

(2) verify the success of the proposed method in frequency alignment quantitatively and qualitatively. 

\subsection{Datasets}
\label{sec:datasets}


\textbf{Real-world face image dataset:} The real images are from CelebA \cite{liu2015deep}, a large-scale image dataset consisting of more than $200k$ real-world celebrity face photos. We randomly sampled $22,000$ images from CelebA as the real image dataset $\mathbb{I}^{+}$. All these images were cropped to a resolution of $128*128*3$ with the face centered and directions aligned.

\noindent \textbf{GAN-forged face dataset:} We selected two powerful and representative GANs, ProGAN and StyleGAN, as the source forgery GAN models to create fake face samples. The two GANs follow the official implementations and are pre-trained with the entire CelebA dataset, allowing them to generate high-fidelity forgery samples. For each GAN, we queried $22,000$ images to construct the fake image dataset $\mathbb{I}^{-}$.

For each class, the $22,000$ images were randomly divided into $20,000$ and $2,000$ as the training set and test set, respectively. The proposed frequency alignment method was developed with the above datasets.

\noindent \textbf{Perturbed forgery images:} We crafted different types of perturbed forgery images based on the $2,000$ ProGAN test images as reference attack samples. We considered three common image processing perturbations \cite{frank2020leveraging, yu2019attributing}, a gradient-based adversarial attack FGSM, and two attacks specific to AI-generated images, GANprintR \cite{neves2020ganprintr} and the state-of-the-art method TR-Net \cite{liu2022making}. The configurations are as follows:

(1) Blurring: images were blurred with a Gaussian filter with a kernel size randomly sampled from $\{3,5,7,9\}$.

(2) Compression: images were JPEG-compressed with a quality factor randomly sampled from $U(10,75)$.

(3) Noising: images were embedded with i.i.d Gaussian noise with a variance randomly sampled from $U(5.0,20.0)$.

(4) FGSM: the adversarial examples were crafted based on the gradient of a vanilla well-trained ResNet detector with two sets of noise amount constraints, i.e., $\epsilon \in \{4/255, 8/255\}$.

(5) GANprintR and TR-Net: we followed the original papers' settings to generate attack samples.





\subsection{Evaluation metrics}
\textbf{Attack performance.} Following previous anti-forensics studies \cite{hussain2021adversarial, neekhara2021adversarial, carlini2020evading, gandhi2020adversarial, fan2021deepfake, barni2019transferability, zhao2020effect}, we evaluate an attack by error rate (ER) and image quality. The ER score is computed as the percentage that test fake samples are mis-classified into 'real' by the detector. Image quality is measured by two widely-used image quality metrics peak signal-to-noise ratio (PSNR) and structural similarity index measure (SSIM). PSNR quantifies the amount of noise that affects the fidelity of an image. SSIM measures the similarity between an original fake image and the corresponding attack sample.

\noindent \textbf{Defense performance.} We report the detection accuracy on fake images (Acc) computed as the percentage that test fake samples are correctly classified to show the performance of a detector. 

\noindent \textbf{Real-referenced Spectral Profile Distance.} To quantify the average frequency discrepancy between ground-truth real images and test (fake) images, we additionally propose a novel metric called Real-referenced Spectral Profile Distance (RSPD), defined as follows:
\begin{equation}
    \mathrm{RSPD}=\frac{1}{N/2}\left(\sum_{k=0}^{N/2-1} \left|\overline{FD^{+}}(r_{k})-\overline{FD^{test}}(r_{k})\right|\right),
\end{equation}
where $\overline{FD^{+}}(r_{k})$ and $\overline{FD^{test}}(r_{k})$ denote the mean 1D spectral profile (computed by Eq. \ref{eq:fd}) averaged over $K$ real images and $K$ test (fake) images, respectively. RSPD measures the average discrepancy between frequency distributions of real and fake images. A lower RSPD score indicates a smaller frequency discrepancy. 

\subsection{Experiment configuration}
Regarding the SMR algorithm, we set the sampling number $K$ to $50$, and the frequency threshold $r_{T}$ to $0.2$ for all experiments. These parameters are chosen based on the trade-off between effectiveness and computational complexity. For the RDC algorithm, we train the auto-encoder $A(\cdot)$ with the $20,000$ CelebA training samples. The batch size is $80$. We use the Adam optimizer \cite{kingma2015adam} with initial learning rates of $1.6e-3$ plus a decay rate of $0.5$ executed at the end of an epoch if the loss stopped decreasing. The loss weight $\lambda$ is $10$. We also use random Gaussian noise, color jitter, and blurring and rotation for data augmentation for training $A(\cdot)$. 

\subsection{Results of the attack implementation}
\subsubsection{Victim detectors}
To demonstrate the transferablity of the attack, we consider a large variety of victim detectors:

\textbf{Normal detectors}: We employ three image-domain detectors based on pixel input, including a ResNet18 and a Xception which are two popular forgery detector backbones, and the GAN fingerprinting model (GF) \cite{yu2019attributing} which learns model fingerprint for detection. We also employ four frequency-domain detectors, one trained with the DCT coefficients (DCT) \cite{frank2020leveraging}, one with 1D spectral profile (1d-SP) \cite{durall2020watch}, as well as the Spatial-Phase Shallow Learning (SPSL)  \cite{liu2021spatial} and the Hierarchical Fine-Grained Image Forgery Detection and Localization (HiFi) \cite{guo2023hierarchical} which combine RGB pixels and frequency spectra for detection. 

\textbf{Specific detectors}: We also consider five detectors with specific design for improving generalization ability and robustness, including the spectral artifacts simulation method (AutoGAN) \cite{zhang2019detecting}, the data augmentation-based method (DA) \cite{wang2020cnn}, the frequency-level adversarial perturbation method (FLAP) \cite{jeong2022frepgan}, the re-synthesis residual method (RSR) \cite{he2021beyond}, and the pretrained vision-language encoder (VLE) \cite{ojha2023towards}.

All detectors are trained with $20,000$ CelebA and $20,000$ ProGAN images and tested with different types of perturbed ProGAN images ($2,000$ per type).  

\subsubsection{Attack performance}
Table \ref{tab:attack_acc} illustrates the performances of eight attack methods in terms of ER and RSPD scores. All detectors except AutoGAN achieve fairly high accuracy in detecting clean ProGAN samples, while their performances degrade when facing attack samples. In the group of normal detectors, our attack evades all detectors, with results comparable to or better than the state-of-the-art attack TR-Net. We also emphasize that the success of FGSM against ResNet18 is not surprising because the attack samples are crafted directly based on the gradient of ResNet18. 

Compared with the normal detector group, the results in the group of specific detectors are more encouraging. Although the effects of almost all attacks are diminished against detectors with strengthened generalization ability and robustness, our attack still leads to relatively high ER scores for all detectors. Moreover, its superiority over other attacks in this group is much more significant than in the normal group. 

This superiority is fully explicable from the frequency perspective. In this group, all the strategies used for enhancing the detectors' generalization ability and robustness can be interpreted as a kind of frequency-domain augmentation, which increases the variety of frequency patterns in the training set and thereby reduces the detector's frequency bias. For example, DA augments the training set with JPEG compression and Gaussian noise, which expands the frequency diversity of the original training samples; FLAP takes one step further by directly generating adversarial perturbations onto the spectra of the original training samples. As a result, one attack will be less effective against the frequency-augmented detectors if it simply modifies the frequency pattern of the original fake samples rather than eliminating the frequency discrepancy between real and fake images, as our method does. The phenomenon also confirms our hypothesis of frequency bias. 

To further verify that the proposed frequency alignment method can reduce high-frequency discrepancies, we also present the RSPD scores, which directly measure the frequency gap between real and fake images, in the last row of Table \ref{tab:attack_acc}. Our attack achieves an RSPD score of $0.22$, which is more than ten times lower than the second-best score $2.36$ obtained by the TR-Net attack, confirming the effectiveness of frequency alignment.

\begin{table*}[htbp]
  \centering
  \caption{The evaluation of the attack implementation in terms of error rate (ER, $\%$, $\uparrow$) and Real-referenced Spectral Profile Distance (RSPD, $\%$, $\downarrow$). A total of Eight attack methods against twelve representative detectors are evaluated. The best result in each row is in bold.}
      \renewcommand\arraystretch{1.2}
  \setlength{\tabcolsep}{1.5mm}{
    \begin{tabular}{l|p{14mm}<{\centering}p{14mm}<{\centering}p{14mm}<{\centering}p{14mm}<{\centering}p{14mm}<{\centering}p{14mm}<{\centering}p{14mm}<{\centering}p{14mm}<{\centering}p{14mm}<{\centering}}
    \toprule
          & Clean & Blurring & Compression &  Noise & FGSM ($\epsilon$=4/255) & FGSM ($\epsilon$=8/255) & GANprintR \cite{neves2020ganprintr} & TR-Net \cite{liu2022making} & FA (ours) \\
        \midrule
    ResNet18 \cite{he2016deep} & 0.08  & 54.65 & 54.15 & 39.17 & 98.61 & \textbf{100.00} & 37.62 & 80.23 & 90.10 \\
    Xception \cite{chollet2017xception} & 0.01  & 46.31 & 41.26 & 50.30 & 78.85 & \textbf{81.66} & 25.12 & 75.50 & 80.31 \\
    GF \cite{yu2019attributing} & 0.11  & 64.20 & 50.91 & 47.37 & 55.55 & 67.00 & 40.99 & \textbf{85.12} & 81.32 \\
    DCT \cite{frank2020leveraging} & 0.06  & 41.53 & 43.00 & 38.46 & 60.01 & 66.31 & 20.13 & 80.01 & \textbf{87.06} \\
    1d-SP \cite{durall2020watch} & 2.63  & 84.99 & 61.51 & 65.03 & 53.21 & 54.47 & 49.52 & 95.98 & \textbf{100.00} \\
    SPSL \cite{liu2021spatial} & 0.06  & 35.17 & 33.85 & 29.60 & 43.60 & 50.12 & 28.88 & \textbf{69.51} & 64.51 \\
    HiFi \cite{guo2023hierarchical} & 0.49  & 54.48 & 47.45 & 44.99 & 64.97 & 69.93 & 33.71 & 81.06 & \textbf{83.88} \\
    \midrule
    AutoGAN \cite{zhang2019detecting} & 18.13 & 20.02 & 31.02 & 30.21 & 40.69 & 45.31 & 75.96 & 82.33 & \textbf{88.16} \\
    DA \cite{wang2020cnn} & 0.03  & 11.36 & 3.63  & 3.21  & 63.70 & 68.83 & 16.34 & 70.03 & \textbf{72.11} \\
    FLAP \cite{jeong2022frepgan} & 1.56  & 4.33  & 16.42 & 10.50 & 40.98 & 35.01 & 11.12 & 62.21 & \textbf{76.36} \\
    RSR \cite{he2021beyond} & 0.01  & 8.16  & 10.66 & 8.19  & 30.12 & 29.63 & 6.89  & 40.79 & \textbf{67.21} \\
    VLE \cite{ojha2023towards} & 0.00  & 18.22  & 30.19 & 15.31  & 20.87 & 27.09 & 9.65  & 22.74 & \textbf{33.09} \\
    \midrule
    \textit{Avg. ER} & 4.93  & 10.97 & 15.43 & 13.03 & 43.87 & 44.70 & 27.58 & 63.84 & \textbf{75.96} \\
    \midrule
    \midrule
    \textit{RSPD (\%) $\downarrow$}  & 4.44  & 9.31  & 4.16  & 13.51  & 8.03  & 12.22  & 5.31  & 2.36  & \textbf{0.22} \\
    \bottomrule
    \end{tabular}}%
  \label{tab:attack_acc}%
\end{table*}

\begin{table*}[htbp]
  \centering
  \renewcommand\arraystretch{1.2}
  \caption{The evaluation of image quality of eight attack methods in terms of the SSIM ($\uparrow$) and PSNR ($\uparrow$) scores. The best result in each row is in bold.}
    \begin{tabular}{l|cccccccc}
    \toprule
          & Blurring & Compression &  Noise & FGSM ($\epsilon$=4/255) & FGSM ($\epsilon$=8/255) & GANprintR & TR-Net & FA (ours) \\
    \midrule
    \textit{PSNR} $\uparrow$ & 28.21 & 33.10 & 30.01 & 31.21 & 30.13 & 27.64 & 35.11 & \textbf{37.91} \\
    \textit{SSIM} $\uparrow$ & 0.714 & 0.886 & 0.766 & 0.812 & 0.760 & 0.901 & \textbf{0.982} & 0.976 \\
    \bottomrule
    \end{tabular}%
  \label{tab:attack_iq}%
\end{table*}%

\subsubsection{Image quality}
\begin{figure*}
    \centering
    \includegraphics[width=0.9\textwidth]{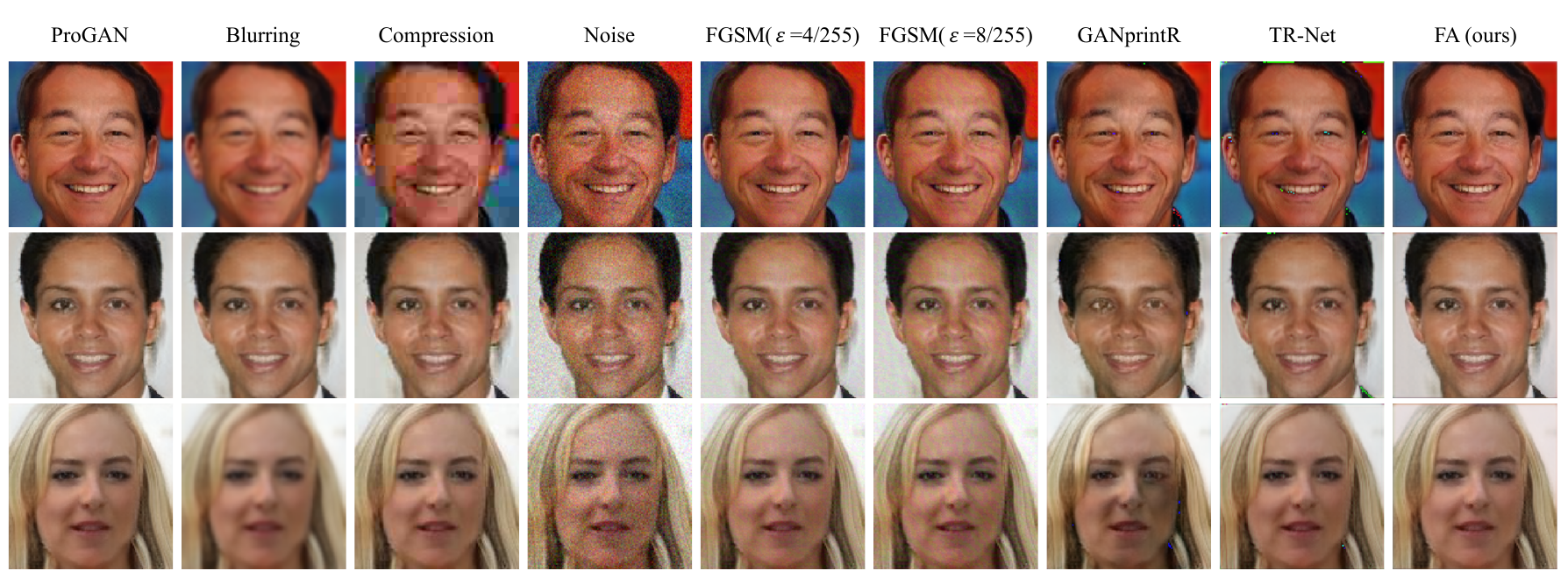}
    \caption{The visualization of three original ProGAN image examples and the corresponding attack samples created by eight attack methods.}
    \label{fig:attack_iq}
\end{figure*}

Another concern about an attack is whether it can maintain the image quality as high as the original fake sample. Table \ref{tab:attack_iq} shows the PSNR and SSIM scores of all attack methods. We can see that, alongside the pronounced attack success, the proposed method achieves the highest PSNR score ($37.91$) and the second-best SSIM score ($0.976$) compared with other attacks. Figure \ref{fig:attack_iq} additionally offers several image examples for explicit visualization. Compared with other attacks, the distortion and noise introduced to the original fake samples by our method are the smallest, which is imperceptible to human eyes. We also emphasize the comparison with TR-Net. Even though TR-Net has a slightly higher SSIM score than our method, we can see in Figure \ref{fig:attack_iq} that it leads to visible point-like noises on the attack samples due to the side effect of adversarial learning discussed in Section \ref{sec:comparison}.

In summary, the results of attack implementation confirm the feasibility of the proposed frequency alignment method as a novel black-box attack. The aligned fake samples are intrinsically closer to real images by removing the frequency discrepancy while maintaining a high visual quality, resulting in great attack transferability across various detectors. 

\subsection{Results of the defense implementation}
Next, we evaluate the effectiveness of the frequency alignment method as a defense strategy for improving a forgery detector's generalization ability and robustness. We evaluate the three different implementation protocols of the frequency alignment method (denoted as FA-P1, FA-P2 and FA-P3, respectively) outlined in Section \ref{sec:def_imple}, plus two baselines, including training with the original dataset (Original) and training with the mixture data augmentation method proposed in \cite{wang2020cnn} (MDA). ResNet18 and Xception are selected as the target detectors. For each detector, we train it from scratch five times, each time with an individual strategy. 

\subsubsection{Generalization}
We first evaluate the generalization ability. The detectors are trained with images from a single GAN and tested with different GANs. Table \ref{tab:defense_gen} shows the detection accuracy of two detectors in different test groups. The right and left sides of the arrow indicate the sources of training samples and test samples, respectively. For example, "P$\rightarrow$S" means training with ProGAN images and testing with StyleGAN images. If the two sides of the arrow are different, it is a cross-GAN test group. 

As shown in Table \ref{tab:defense_gen}, both of the original ResNet18 and Xception can achieve high detection accuracy in the intra-dataset tests P $\rightarrow$ P and S $\rightarrow$ S, even without any defense. However, when generalized to the cross-GAN tests, their performances drop considerably. For example, the Acc score of the ResNet18 trained with the ProGAN images decreases from $99.21\%$ in the P $\rightarrow$ P group to $12.31\%$ in the P $\rightarrow$ S group. The results of the original detectors indicate that a normal DNN can excessively learn the difference between real images and images generated by a specific GAN, which may lead to overfitting. By comparison, after implementing a defense strategy, the generalization abilities of both detectors get highly improved in all cross-GAN tests. 

Among the four defense strategies, FA-P2 and FA-P3 are much more effective than MDA and FA-P1 in enhancing the cross-GAN generalization ability while maintaining the intra-dataset detection accuracy. One possible reason is that MDA and FA-P1 are both based on data augmentation, which reduces the frequency bias of the detector by improving the frequency diversity in the training set only. In contrast, FA-P2 and FA-P3 implement frequency alignment as a pre-processing module for both the training and test samples. It can pull all samples to the same distribution in the frequency domain prior to detection, so as to eliminate the frequency bias in detection.

\begin{table}[htbp]
  \centering
  \caption{The evaluation of generalization ability in the defense implementation in terms of detection accuracy (Acc, $\%$, $\uparrow$). A total of five defense protocols are evaluated in four test groups. The best result in each column is in bold.}
  \renewcommand\arraystretch{1.2}
  \setlength{\tabcolsep}{2.2mm}{
    \begin{tabular}{ll|cc|cc}
    \toprule
    \multicolumn{2}{c|}{\textit{Protocols}} & \multicolumn{1}{l}{P$\rightarrow$P} & \multicolumn{1}{l|}{S$\rightarrow$S} & \multicolumn{1}{l}{P$\rightarrow$S} & \multicolumn{1}{l}{S$\rightarrow$P} \\
    \midrule
    \multirow{5}[2]{*}{ResNet18} & Original & 99.92  & 99.19  & 12.31  & 31.36  \\
          & MDA \cite{wang2020cnn}  & 98.78  & 98.12  & 59.21  & 67.23  \\
          & FA-P1 (ours)  & 98.91  & 98.22  & 61.41  & 68.45 \\
          & FA-P2 (ours)  & 99.61  & \textbf{100.00} & 81.13  & \textbf{86.10} \\ 
          & FA-P3 (ours) & \textbf{100.00} & \textbf{100.00} & \textbf{83.20} & 85.21  \\
    \midrule
    \midrule
    \multirow{5}[2]{*}{Xception} & Original & \textbf{100.00} & 99.93  & 32.18  & 40.77  \\
          & MDA \cite{wang2020cnn}  & 98.65  & 98.55  & 54.13  & 73.09  \\
          & FA-P1 (ours) & 98.81  & 98.23  & 60.20  & 72.67  \\ 
          & FA-P2 (ours) & 99.36  & \textbf{100.00} & 81.32  & 82.69  \\
          & FA-P3 (ours) & \textbf{100.00} & \textbf{100.00} & \textbf{87.02} & \textbf{84.03} \\
    \bottomrule
    \end{tabular}}%
  \label{tab:defense_gen}%
\end{table}%

\subsubsection{Robustness}
We next evaluate the robustness against different attacks. The detectors are trained with the clean ProGAN images and tested with the seven types of attack samples described in Section \ref{sec:datasets}, plus the proposed FA attack samples. Note that our FA-based defenses become white-box only when detecting the FA attack samples, while other settings are all black-box. Table \ref{tab:defense_rob} demonstrates the results in terms of detection accuracy. The bare DNN-based detectors suffering from severe frequency bias are vulnerable to various perturbation attacks, especially the adversarial attack FGSM, resulting in low detection accuracy scores. When being strengthened by defenses that can mitigate the frequency bias, the detectors become more reliable in classifying attack samples.

Regarding different defense strategies, similar to the results of generalization ability in Table \ref{tab:defense_gen}, FA-P2 and FA-P3 are generally more effective than MDA and FA-P1 when dealing with all attack types. Note that MDA also performs well in the Compression and Noise groups. This is because MDA uses compression and noise for data augmentation \cite{wang2020cnn}; thus, it becomes a de facto white-box defense in the two groups. In comparison, the proposed FA-P2 and FA-P3 are more practical since they work evenly well for different attacks without knowing the attack setting. 

In summary, the results of defense implementation showcase the potential of the proposed frequency alignment method being a universal strategy for improving generalization and robustness of a forgery detector. It is effective for various unknown forgery patterns and compatible with different detectors. 

\begin{table*}[htp]
  \centering
  \caption{The evaluation of the robustness of different defense implementations in terms of detection accuracy (Acc, $\%$, $\uparrow$). A total of five defense protocols are evaluated against eight attack methods, including our proposed FA attack. The best result in each column is in bold. *Note that our FA-based defenses become white-box only when detecting the FA attack samples, while other settings are all black-box.}
  \renewcommand\arraystretch{1.2}
  \setlength{\tabcolsep}{1.5mm}{
    \begin{tabular}{cl|p{14mm}<{\centering}p{14mm}<{\centering}p{14mm}<{\centering}p{14mm}<{\centering}p{14mm}<{\centering}p{14mm}<{\centering}p{14mm}<{\centering}p{14mm}
    <{\centering}}
    \toprule
    \multicolumn{2}{c|}{\textit{Protocols}} & Blurring & Compression &  Noise & FGSM ($\epsilon$=4/255) & FGSM ($\epsilon$=8/255) & GANprintR \cite{neves2020ganprintr} & TR-Net \cite{liu2022making} & FA (ours)*\\
    \midrule
    \multirow{5}[2]{*}{ResNet18} & Original & 45.35 & 45.85 & 60.83 & 1.39  & 0.00  & 62.38 & 19.77 & 9.90\\
          & MDA \cite{wang2020cnn}  & 76.79  & 90.11  & \textbf{93.68} & 49.88  & 49.91  & 70.43  & 31.69  & 27.89\\
          & FA-P1 (ours) & 70.46  & 74.70  & 77.01  & 65.39  & 60.93  & 71.12  & 58.97 & 95.81\\
          & FA-P2 (ours) & 86.71  & 87.27  & 87.45  & \textbf{85.04} & 84.99  & \textbf{91.78} & 79.59  & 96.41\\
          & FA-P3 (ours) & \textbf{88.03} & \textbf{90.89} & 90.39  & 81.89  & \textbf{85.64} & 90.55  & \textbf{80.80} & \textbf{98.97}\\
    \midrule
    \midrule
    \multirow{5}[2]{*}{Xception} & Original & 53.69 & 58.74 & 49.70 & 21.15 & 18.34 & 74.88 & 24.50 & 20.69\\
          & MDA \cite{wang2020cnn}  & 80.50  & 90.88  & \textbf{91.91} & 43.13  & 46.74  & 74.10  & 51.95  & 35.66\\
          & FA-P1 (ours) & 68.80  & 67.56  & 73.23  & 70.39  & 69.46  & 72.88  & 58.00 & 94.88 \\
          & FA-P2 (ours) & \textbf{89.98} & 89.69  & 88.89  & 83.72  & 82.97  & 89.41  & 75.48 & 94.30\\
          & FA-P3 (ours) & 88.22  & \textbf{91.37} & 89.74  & \textbf{87.10} & \textbf{86.19} & \textbf{90.78} & \textbf{83.36} & \textbf{98.19}\\
    \bottomrule
    \end{tabular}}%
  \label{tab:defense_rob}%
\end{table*}%

\subsection{The effect of frequency alignment} 

\subsubsection{Visualizations}
The key effect of the proposed frequency alignment method is that it can align the frequency pattern of an arbitrary type of fake image to real images, eliminating the frequency discrepancy and making fake images intrinsically closer to real ones. We now provide some visualizations to confirm the effect. 

\begin{figure}
    \centering
    \includegraphics[width=0.48\textwidth]{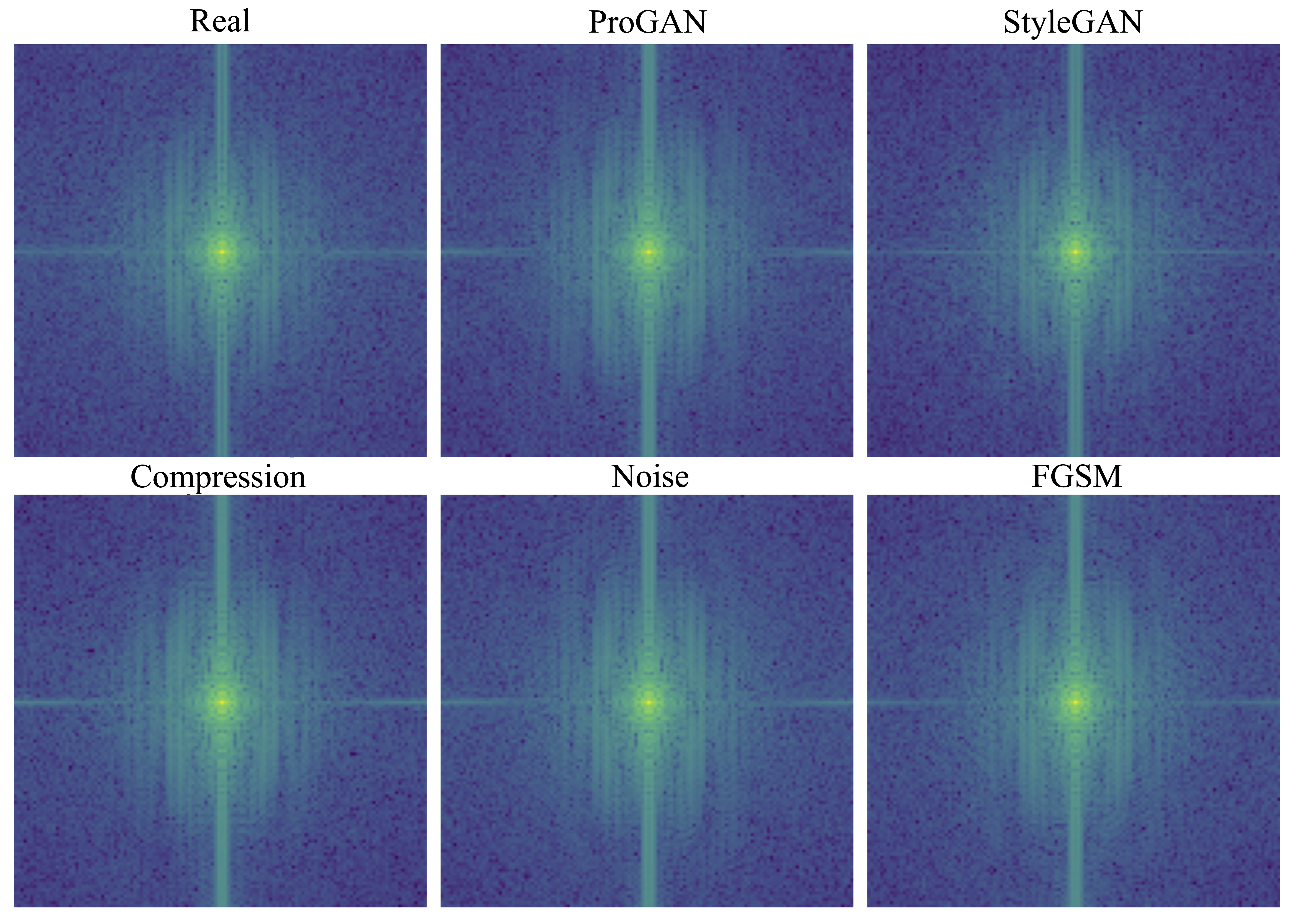}
    \caption{The average DFT spectra of real images and different types of frequency-aligned AI-generated images and images perturbed from clean ProGAN images. A visual comparison can be made with Figure \ref{fig:spectral_dis}, where the higher-frequency artifacts in the original spectra in Figure \ref{fig:spectral_dis} has been removed thoroughly.}
    \label{fig:spectral_dis_after}
\end{figure}

\begin{figure}
    \centering
    \includegraphics[width=0.48\textwidth]{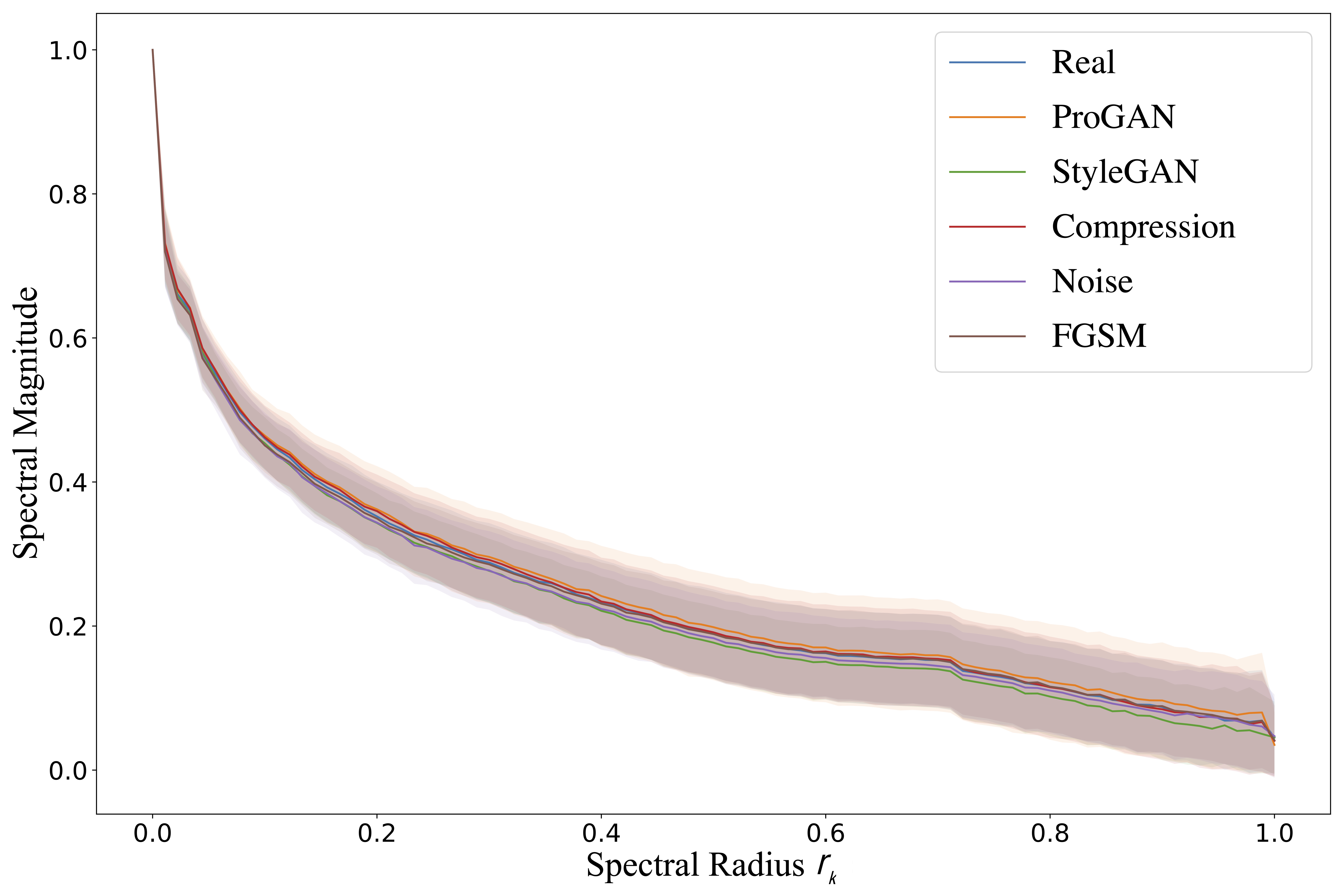}
    \caption{Visualization of the spectral profiles of real images and different types of frequency-aligned fake images. The distributional gaps are removed compared with Figure \ref{fig:distributional_dis}.}
    \label{fig:distributional_dis_after}
\end{figure}

First, we visualize the average DFT spectra and frequency distributions of real and frequency-aligned fake images. We select the forgery types covered in Section \ref{sec:fafd} for a straightforward comparison. Figure \ref{fig:spectral_dis_after} and Figure \ref{fig:distributional_dis_after} display the average DFT spectra and the frequency distributions, respectively. From Figure \ref{fig:spectral_dis_after}, we can see that after frequency alignment, different types of fake images all manifest a spectral pattern similar to the pattern of real images, compared with Figure \ref{fig:spectral_dis}. Particularly, the spectral artifacts shown in Figure \ref{fig:spectral_dis} has been suppressed thoroughly. This effect can be further confirmed in the frequency distributions shown in Figure \ref{fig:distributional_dis}. The frequency distributions of the aligned fake images now become consistent with real images, in contrast to the substantial distributional gaps exhibited in Figure \ref{fig:distributional_dis}. The results confirm the success of frequency alignment as well as the broad applicability of the proposed method to various forgery types.

\begin{figure}
    \centering
    \includegraphics[width=0.45\textwidth]{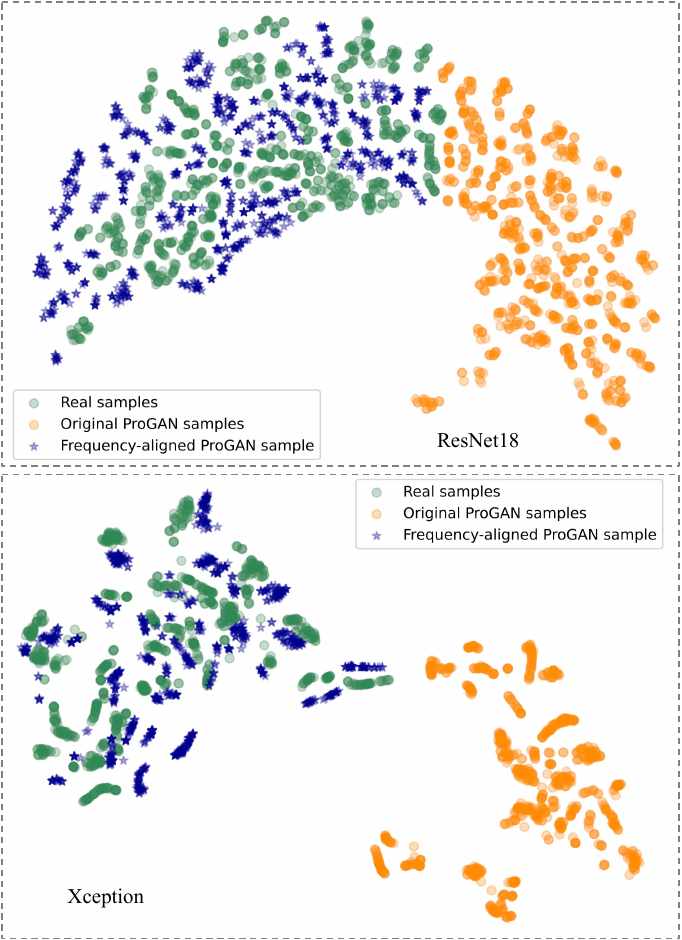}
    \caption{The features of real, original ProGAN, and frequency-aligned ProGAN images in the latent spaces of ResNet18 and Xception detectors.}
    \label{fig:tsne}
\end{figure}

To further demonstrate that the frequency alignment method enables fake images to be truly closer to real images, we visualize the changes in DNN detectors' latent feature space caused by frequency alignment. Figure \ref{fig:tsne} shows the features of real, original ProGAN, and frequency-aligned ProGAN images extracted from the last convolutional layer of the ResNet18 and Xception detectors. Features are clustered into a two-dimensional space by T-SNE \cite{van2008visualizing} for visualization. As shown in the figure, the features of frequency-aligned ProGAN images are entangled with those of real images, while far away from the features of original ProGAN images. The results also explain why the frequency-aligned fake images can be used as attack samples to evade detection.  

\begin{table*}[htbp]
  \centering
  \caption{The PSNR, SSIM and RSPD scores of different frequency alignment methods.}
    \renewcommand\arraystretch{1.2}
    \setlength{\tabcolsep}{2mm}{
    \begin{tabular}{l|cccccccc}
    \toprule
          & BLF ($r_0=0.2$) & BLF ($r_0=0.5$) & BLF ($r_0=0.8$) & FA-AL & FA-SMR & FA-RDC & FA-Final \\
    \midrule
    \textit{PSNR} $\uparrow$ & 28.33  & 29.39  & 37.02  & 36.71  & 32.17  & 36.01  & \textbf{37.91}  \\
    \textit{SSIM} $\uparrow$ & 0.721  & 0.749  & 0.961  & 0.959  & 0.903  & 0.955  & \textbf{0.976}  \\
    \midrule
    \midrule
    \textit{RSPD} $\downarrow$ & 8.17  & 5.90  & 4.92  & 4.09  & 1.08  & 0.62  & \textbf{0.22}  \\
    \bottomrule
    \end{tabular}}%
  \label{tab:ablation}%
\end{table*}%

\begin{figure*}
    \centering
    \includegraphics[width=0.9\textwidth]{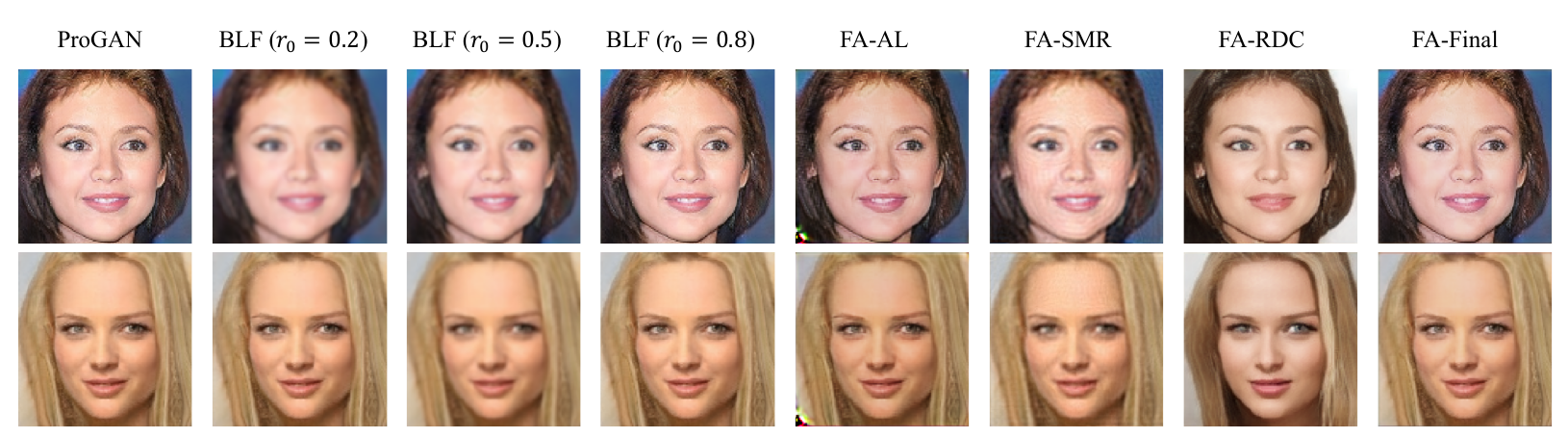}
    \caption{Examples of the original ProGAN images and the versions modified by different frequency alignment methods.}
    \label{fig:ablation}
\end{figure*}
\subsubsection{Comparison and ablation study}
We compare the proposed frequency alignment method to other possible methods outlined in Section \ref{sec:comparison}. For the low-pass filter method, we test a Butterworth Low-pass Filter (BLF) with varying cut-off bands $r_0$. Regarding the adversarial learning method, we redesign our RDC algorithm by adding a frequency discriminator $D$ complement to the auto-encoder $A$ and train them adversarially, denoted as FA-AL. The frequency regularization methods require modifying the source GAN, which is not applicable to this evaluation. We also provide an ablation study that compares the individual SMR (FA-SMR) and RDC (FA-RDC) algorithms to the final combined version (FA-Final).

We assess these methods based on the frequency alignment and image quality trade-offs. Table \ref{tab:ablation} shows the results in terms of PSNR, SSIM and RSPD scores. Figure \ref{fig:ablation} depicts some examples of the original ProGAN images and the versions modified by the above methods. Compared with other learning-based methods, BLF with a fixed cut-off band is inflexible and fails to align the frequency accurately. Moreover, when the cut-off band is small, the image quality is readily degraded. Adopting an adversarial discriminator like in FA-AL raises the SSIM and RSPD scores, but falls short in  frequency alignment. This is because, as discussed in Section \ref{sec:comparison}, the discriminator will also suffer from the frequency bias. In addition, despite the high SSIM and RSPD scores, the adversarial learning method will leave particular visual speckles on the output images, such as in the lower-left corner of the image, as shown in Figure \ref{fig:ablation}. Comparatively, the frequency alignment effect achieved by the individual SMR or RDC algorithm surpasses the baselines remarkably. However, the image quality remains a concern: FA-SMR will produce conspicuous visual artifacts, whilst FA-RDC will cause color and local texture distortions. The combined version FA-Final, obtains the highest PSNR, SSIM and RSPD scores, indicating the best trade-off between frequency alignment and image quality. 

\section{Conclusion}
Machine-generated image forgeries continue to threaten digital ecosystems and personal security. High-fidelity, fully automated GAN-based forgeries amplify this risk, underscoring the need for reliable detection. This work provides a principled, frequency-based explanation for two core reliability factors—generalization and robustness, and reveals that specific spectral discrepancies between real and fake images induce a DNN frequency bias that jointly degrades both properties.

To address this, we introduce a two-step frequency alignment method that reduces spectral gaps between real and forged images. The approach is simple to implement and yields dual benefits: as a strong black-box attack, frequency-aligned forgeries shift toward the real-image distribution and evade detectors; as a universal defense, alignment decreases detector frequency bias, improving generalization and robustness without architectural changes. We present corresponding attack and defense implementations and demonstrate their effectiveness across diverse evaluations, alongside clear evidence of the alignment effect.

This study provides a foundation for enhancing the reliability of deep forgery detectors. Future directions include extending the approach to additional image domains and modalities (e.g., medical or remote sensing imagery, video and audio deepfakes), and exploring curriculum or self-supervised schemes that further mitigate frequency bias in open world.




\bibliographystyle{IEEEtran}
\bibliography{IEEEabrv,./ref.bib}

\newpage

\begin{IEEEbiography}[{\includegraphics[width=1in,height=1.25in,clip,keepaspectratio]{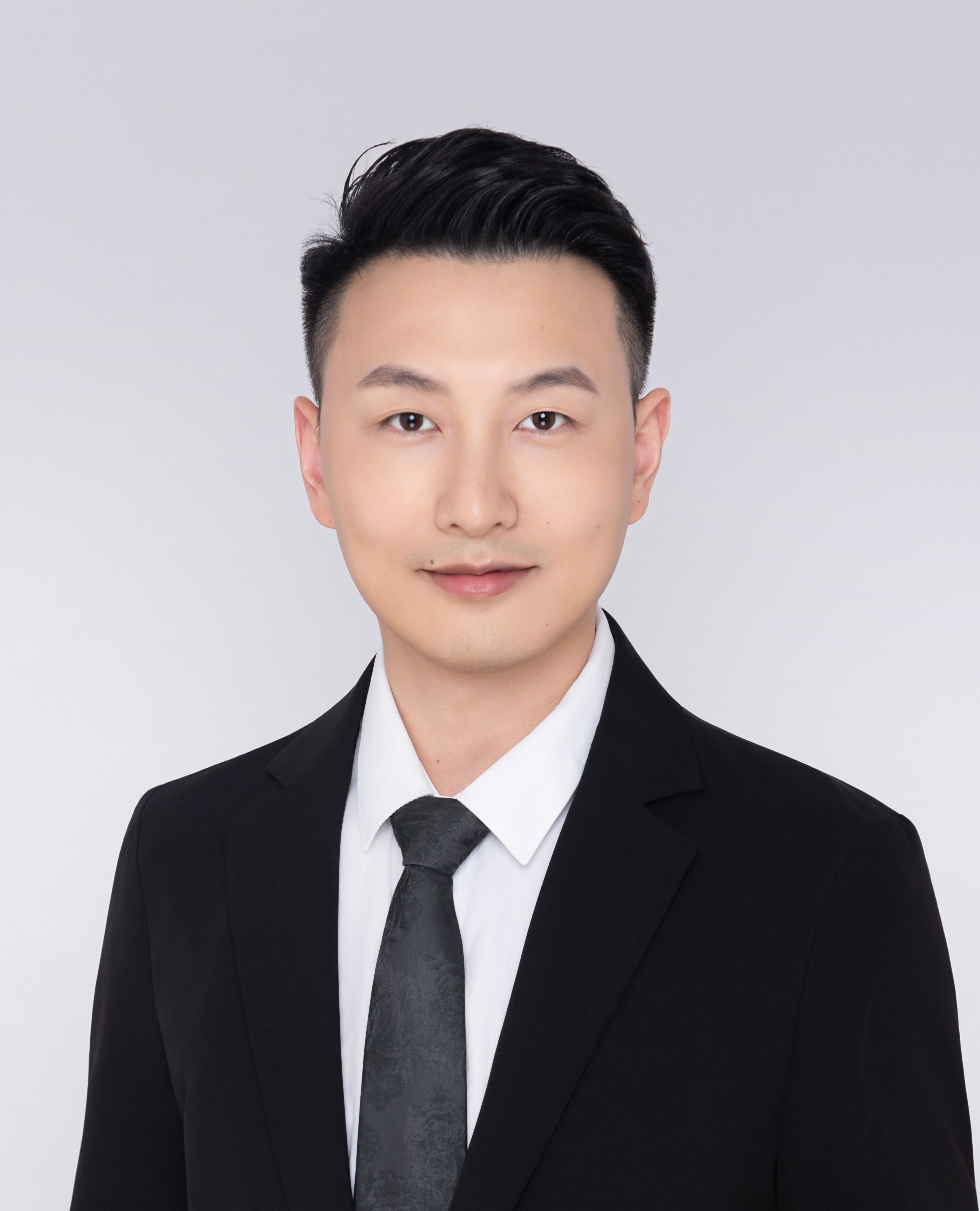}}]{Chi Liu} is an Assistant Professor at the Faculty of Data Science, City University of Macau, Macao SAR, China. Prior to this, he was a research fellow at the Faculty of Information Technology, Monash University. He received a Ph.D. from the School of Computer Science, University of Technology Sydney, Australia, an M.Sc. from the China Ship Research and Development Academy, China, and a B.Sc. from Wuhan University, China. His research interests include AIGC safety, trustworthy and responsible AI, etc.
\end{IEEEbiography}
\vspace{-10pt} 
\begin{IEEEbiography}[{\includegraphics[width=1in,height=1.25in,clip,keepaspectratio]{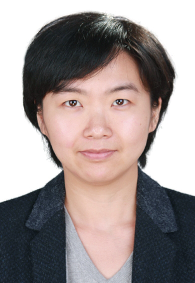}}] {Tianqing Zhu}
is currently a Professor with the Faculty of Data Science, City University of Macau, Macau, SAR, China. Prior to this, she was an Associate Professor with the University of Technology Sydney, Ultimo, NSW, Australia, and a Lecturer with the School of Information Technology, Deakin University, Australia. She received the B.Eng. and M.Eng. degrees from Wuhan University, Wuhan, China, in 2000 and 2004, respectively, and the Ph.D. degree from Deakin University, Sydney, Australia, in 2014. Her research interests include privacy-preserving, data mining, and AI security.
\end{IEEEbiography}
\vspace{-10pt} 
\begin{IEEEbiography}[{\includegraphics[width=1in,height=1.25in,clip,keepaspectratio]{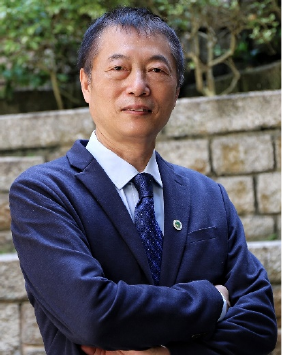}}]{Wanlei Zhou}
Professor Wanlei Zhou (Senior member, IEEE) is currently the Vice Rector (Academic Affairs) and Dean of Institute of Data Science, City University of Macau, Macao SAR, China. He received the PhD degree in Computer Science and Engineering from The Australian National University, Canberra, Australia, in 1991. He also received a DSc degree (a higher Doctorate degree) from Deakin University in 2002. Before joining City University of Macau, Professor Zhou held various positions including the Head of School of Computer Science in University of Technology Sydney, Australia, the Alfred Deakin Professor, Chair of Information Technology, Associate Dean, and Head of School of Information Technology in Deakin University, Australia. His main research interests include security, privacy, and distributed computing. He has published more than 400 papers in refereed international journals and refereed international conference proceedings, including many articles in IEEE transactions and journals.
\end{IEEEbiography}
\vspace{-10pt} 
\begin{IEEEbiography}[{\includegraphics[width=1in,height=1.25in,clip,keepaspectratio]{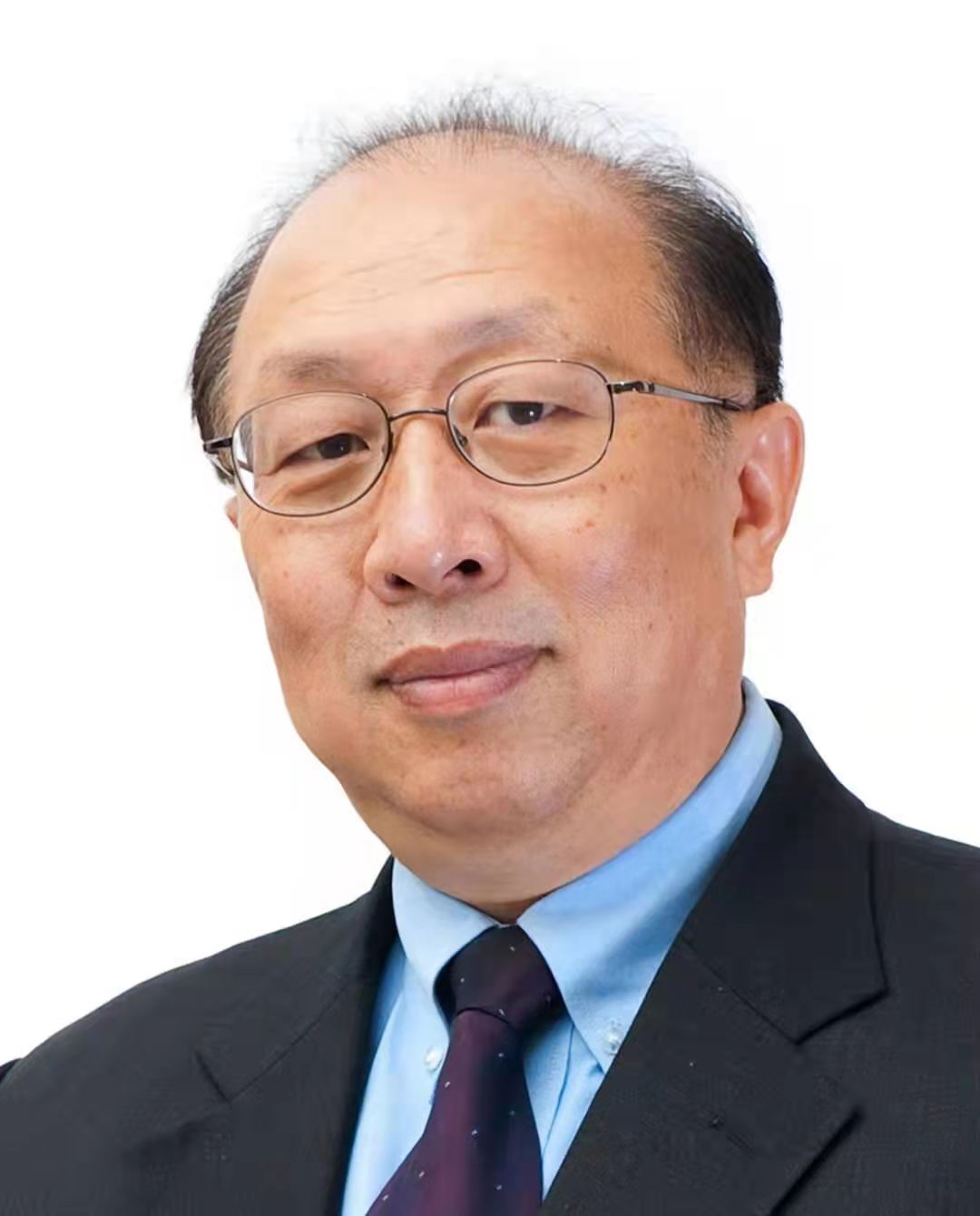}}]{Wei Zhao} as an IEEE fellow, has served important leadership roles in academic including the eighth Rector of the University of Macau, the Chief Research Officer (i.e., Vice President for Research) at the American University of Sharjah, the Chair of Academic Council at CAS Shenzhen Institute of Advanced Technology, the Dean of Science at Rensselaer Polytechnic Institute, the Director for the Division of Computer and Network Systems in the U.S. National Science Foundation, and the Senior Associate Vice President for Research at Texas A\&M University. He was also on faculty in Shaanxi Normal University, Amherst College, Adelaide University, and Shenzhen Institute of Advanced Technologies. Professor Zhao completed his undergraduate studies in physics at Shaanxi Normal University, China, in 1977, and received his MSc and PhD degrees in Computer and Information Sciences at the University of Massachusetts at Amherst in 1983 and 1986, respectively.
\end{IEEEbiography}

\vfill

\end{document}